\documentclass{emulateapj}
\usepackage{amsmath, amssymb}
\usepackage{natbib}
\usepackage{lscape}
\shorttitle{}
\shortauthors{Graves et al.}

\begin{document}

\title{Dissecting the Red Sequence---II. Star Formation Histories of
  Early-Type Galaxies Throughout the Fundamental Plane}

\author{Genevieve J. Graves\altaffilmark{1},
S. M. Faber\altaffilmark{1}, \& Ricardo P. Schiavon\altaffilmark{2}}

\altaffiltext{1}{UCO/Lick Observatory, Department of Astronomy and
  Astrophysics, University of California, Santa Cruz, CA 95064}
\altaffiltext{2}{Gemini Observatory, 670 N. A'ohoku Place, Hilo, HI 96720}

\keywords{galaxies: abundances, galaxies: elliptical and lenticular}

\begin{abstract}

This analysis uses spectra of $\sim$16,000 nearby SDSS quiescent
galaxies to track variations in galaxy star formation histories along
and perpendicular to the Fundamental Plane (FP).  We sort galaxies by
their FP properties ($\sigma$, $R_e$, and $I_e$) and construct high
$S/N$ mean galaxy spectra that span the breadth and thickness of the
FP.  From these spectra, we determine mean luminosity-weighted ages,
[Fe/H], [Mg/H], and [Mg/Fe] based on single stellar population models
using the method described in \citet{graves08}.  In agreement with
previous work, the star formation histories of early-type galaxies are
found to form a two-parameter family.  The major trend is that mean
age, [Fe/H], [Mg/H], and [Mg/Fe] all increase with $\sigma$.  However,
no stellar population property shows any dependence on $R_e$ at fixed
$\sigma$, suggesting that $\sigma$ and not dynamical mass ($M_{dyn}
\varpropto \sigma^2 R_e$) is the better predictor of past star
formation history.  In addition to the main trend with $\sigma$,
galaxies also show a range of population properties at fixed $\sigma$
that are strongly correlated with surface brightness residuals from
the FP ($\Delta \log I_e$), such that higher surface brightness
galaxies have younger mean ages, higher [Fe/H], higher [Mg/H], and
lower [Mg/Fe] than lower-surface brightness galaxies.  These latter
trends are a major new constraint on star-formation histories.
\end{abstract}

\section{Introduction}\label{introduction}

Early-type galaxies are known to comprise a two-parameter family in
galaxy structure.  In the three-dimensional (3-D) parameter space of
central stellar velocity dispersion ($\sigma$), galaxy effective
radius ($R_e$), and mean central surface brightness ($I_e$), early
type galaxies occupy a 2-D plane, known as the Fundamental Plane (FP;
\citealt{faber87,dressler87,djorgovski87}).  The FP is relatively
constant throughout the local universe \citep{jorgensen96}.  There may
be some variation with environment \citep{bernardi03c}, consistent
with the FP in dense environments being offset toward lower $I_e$ than
the FP in the field.

There is clear evolution in the FP with redshift, which is often
interpreted as evolution in the stellar mass-to-light ratio ($M/L$).
For field galaxies, most authors find consistent results, with $M/L$
evolution of $d \log (M/L_B) / dz \sim -0.7$ \citep[and references
therein]{treu05}.  \citet{treu05} also find increased scatter in the
FP at higher redshift, with most evolution seen in the low-mass end;
while the high-mass FP appears nearly unchanged since $z \sim 1$ other
than expected passive evolution in $M/L$, low-mass galaxies are still
``settling'' onto the FP since $z \sim 1$ \citep{vanderwel04,
treu05-apjl, treu05, vanderwel05}.

The local FP should then be expected to include at least some galaxies
that are relatively recent arrivals.  This is supported by the results
of \citet{forbes98} and \citet{terlevich02}, who find that residuals
from the FP correlate with age such that galaxies offset to higher
(lower) surface brightness have younger (older) ages than those
occupying the FP.  The thickness of the FP may therefore be an age
sequence, with the most recent arrivals lying at higher surface
brightness than the midplane.  This thickness may reveal a (narrow)
third dimension to the structural parameters of early-type galaxies.

In contrast to the 2-D structural space occupied by early-type
galaxies, most studies of their stellar populations have stressed only
a 1-D space.  Many previous authors have presented correlations
between detailed stellar population properties and various measures of
galaxy ``size'', including $\sigma$ (e.g., \citealt{trager00b,
thomas05, nelan05, smith07}), stellar mass $M_*$ (e.g.,
\citealt{gallazzi05}), and luminosity ($L$) (e.g.,
\citealt{kuntschner98,terlevich02}).  However, considerable evidence
suggests that early-type galaxy star formation histories in fact
comprise a two parameter family.  \citet{worthey95} demonstrated an
anti-correlation between age and metallicity among galaxies with
similar $\sigma$ and showed that it helped explain the narrowness of
the color-$\sigma$ and FP relations.  \citet{trager00b} quantified
this anti-correlation, which they termed the ``metallicity
hyperplane.''  The anti-correlation has been corroborated recently by
\citet{smith07-iaus245}.  \citet{thomas05} also found a range of ages
for galaxies at fixed $\sigma$ although with no associated metallicity
variation.

This is the second paper in a four-part series which demonstrates
conclusively that early-type galaxies do span a 2-D space in stellar
population properties, implying that their star formation histories
must also occupy a two-parameter space.  The goal of this series is to
connect stellar population properties of early-type galaxies to their
structural parameters in order to understand how galaxy structural
evolution and mass assembly are related to star formation.  In
\citet[hereafter Paper I]{graves08_paperI}, we explored the
variations in stellar populations along the color-magnitude relation
of early-type galaxies and showed that galaxy stellar populations not
only vary systematically with $\sigma$, but also vary with residuals
from the $\sigma$--$L$ relation.

In this paper, we continue that approach but map out stellar
populations along and through the FP rather than the $\sigma$--$L$
relation.  The result is to parameterize the 2-D space of stellar
population variables in terms of the FP variables $\sigma$, $R_e$, and
$I_e$.  Briefly looking foward, Paper III presents a clear
parameterization of the 2-D family of early-type galaxy star formation
histories and interprets the observed stellar population results as
differences in the onset epoch and duration of star formation in
galaxies.  Paper IV compares various stellar ($M_*/L$) and dynamical
($M_{dyn}/L$) mass-to-light ratios for the sample galaxies and
demonstrates that the observed stellar population effects are not
adequate to explain the variation in $M_{dyn}/L$ along or
perpendicular to the FP.  

The paper is organized as follows.  Section \ref{sample} briefly
reviews the selection criteria used to identify the early-type galaxy
sample for this series of papers.  In \S\ref{deconstructing_fp}, we
divide up the FP into bins of similar galaxies, stack the spectra to
obtain high $S/N$ mean spectra for each point in FP-space, and compare
the spectra to models in order to determine the mean stellar
population properties along and through the FP.  Section
\ref{stellar_pops} presents these results.  A major new result is that
stellar populations, and therefore galaxy star formation histories,
appear to be independent of $R_e$, at fixed $\sigma$.  Section
\ref{discussion} briefly discusses some possible explanations for why
galaxy star formation histories do not depend on $R_e$ at fixed
$\sigma$.  Finally, \S\ref{conclusions} summarizes our conclusions.

\section{Sample Selection}\label{sample}

The sample of galaxies used in this work is identical to that of Paper
I.  A brief summary of the sample selection is included here.  The
sample consists of $\sim$16,000 galaxies from the Sloan Digital Sky
Survey (SDSS; \citealt{york00}) spectroscopic Main Galaxy Survey
\citep{strauss02}) Data Release 6 (DR6; \citealt{adelman-mccarthy08}).
Galaxies are chosen from a limited redshift range ($0.04 < z < 0.08$)
and are selected to be quiescent galaxies based on the following
criteria:
\begin{list}{}{}
\item[1.]{Galaxies have no detected emission in either the H$\alpha$ or
  [O\textsc{ii}]$\lambda3727$ lines.}
\item[2.]{Galaxies have centrally concentrated light profiles, as
  determined by the ratio of the Petrosian radii that enclose 90\% and
  50\% of the total galaxy light ($R_{90}/R_{50} > 2.5$ in the
  $i$-band).}
\item[3.]{The likelihood of a de Vaucouleurs light profile fit is at
  least 1.03 times larger than the likelihoood of an exponential light
  profile fit.}
\end{list}
Paper I showed that, although no explicit color selection has been
applied, these criteria define a sample of galaxies that populate
the red sequence in the color-magnitude diagram, with very few color
outliers (see Figure 1 of Paper I).  Remaining color outliers are
excluded from this analysis, as described in
\S\ref{deconstructing_fp}.  

By requiring the sample galaxies to have no detectable emission, we
remove four types of galaxies from the sample: actively star-forming
galaxies, Seyfert hosts, low ionization nuclear emission-line region
(LINER) hosts, and so-called ``transition objects'' (TOs) whose
spectra contain emission from both star-formation and active galactic
nuclei (AGN).  In galaxies with ongoing active star-formation
(star-forming galaxies and TOs), a small fraction of very young stars
contribute disproportionately to the integrated light of the galaxy.
These galaxies will therefore have significantly lower $M_*/L$ than
quiescent galaxies of comparable mass, which will change their
location in FP space.  These galaxies constitute about 11\% of
morphologically-selected early type galaxies \citep{schawinski07}.
Excluding these galaxies removes the ``youngest'' early type galaxies
from the sample, leaving a sample of galaxies whose $M/L$ values and
positions in FP space are not strongly biased by a small
sub-population of very young stars.

Removing galaxies with emission due to AGN activity (i.e., Seyferts
and LINERs) is a less obvious choice, as these galaxies likely lack
the very young stars that bias $M/L$ estimates in actively
star-forming galaxies.  However, host galaxies of strong AGN have
stellar light profiles intermediate between early-type and late-type
galaxies \citep{kauffmann03}, as indicated by values of
$R_{90}/R_{50}$.  Furthermore, there is evidence that these galaxies
have systematically different stellar populations from quiescent early
type galaxies of the same mass, with both Seyferts \citep{kauffmann03}
and LINERs \citep{graves07} showing younger stellar population ages.
In light of these differences, we defer a FP analysis of these
galaxies to future work.

Defined in this way, the galaxy sample presented here contains
bulge-dominated galaxies, some of which harbor significant disk
components.  Figure 2 in Paper I shows a selection of the sample
galaxies illustrating this.  Classical bulges of disk galaxies are
known to lie on the FP defined by early-type galaxies
\citep{fisher08,kormendy08} and are therefore relevant to this study.
Because the SDSS images are fairly low-resolution, high quality
automated bulge-disk decompositions are challenging.  Assessing the
quality of the decompositions and interpreting the output is a
complicated topic and beyond the scope of the analysis presented here.
We are engaged in ongoing work to asses and compare automated
bulge-disk compositions with by-eye morphologies (Cheng et al., in
preparation).  We then intend to revisit the topic of morphological
variation among quiescent galaxies and its effect on the FP and
stellar population.

The FP explored in this work is defined by the galaxy central velocity
dispersion ($\sigma$), galaxy effective radius ($R_e$), and galaxy
effective surface brightness ($I_e$).  Following
\citet{jorgensen96,jorgensen06}, values of $R_e$ are derived using
$r^{1/4}$ de Vaucouleurs fits to the galaxy light profiles (i.e.,
\citealt{sersic68} fits with $n=4$).  It is likely that not all sample
galaxies will have true S{\'e}rsic $n=4$ profiles and that this will
affect the derived FP correlations.  However, it is not clear that a
physically meaningful comparison can be made between values of $R_e$
derived from fitting profiles with different shapes.  A separate study
of the FP relationships presented here as a function of S{\'e}rsic $n$
will be included in the FP morphology-dependence study discussed
above.

In this work, velocity dispersions are taken from the SDSS DR6 catalog
and corrected to a ``central'' $\frac{1}{8} R_e$ aperture (see Paper I
for details).  Effective radii are from $r$-band de Vaucouleurs fits
to the galaxy light profiles, converted to physical units using
standard $\Lambda$CDM cosmology with $\Omega_{\Lambda}=0.7$,
$\Omega_M=0.3$, and $h_0 = 0.70$.  Surface brightnesses are computed
in the Johnson $V$ band as $I_e = L_V / 2 \pi R_e^2$, where $L_V$ is
the total $V$ band luminosity of the galaxy from a de Vaucouleurss fit
to the light profile.  The K-correction from observed SDSS $ugriz$
photometry to rest-frame $V$-band is performed using the IDL code
$kcorrect$ v4.1.4 \citep{blanton03-kcorrect}.  The de Vaucouleurs
radii and luminosities are corrected for known problems with
sky-subtraction around bright galaxies in the SDSS photometric
pipeline, as described in Paper I.  Luminosities are also corrected
for Galactic extinction using the extinction values from the SDSS
photometric pipeline.

The data used to identify the sample and to compute $\sigma$, $R_e$,
and $I_e$ for each galaxy include the following parameters from the
NYU Value-Added Catalog \citep{blanton05-vagc} version of the SDSS DR4
\citep{adelman-mccarthy06}: redshift ($z$), de Vaucouleurss photometry
in the $ugriz$ bands, and Petrosian radii ($R_{90}$ and $R_{50}$).
These are supplemented with the following parameters from the SDSS DR6
Catalog Archive Server\footnote{http://cas.sdss.org/dr6/en}: velocity
dispersion as measured in the SDSS $3''$ fiber ($\sigma_{fib}$),
$r$-band de Vaucouleurs radius ($r_{deV}$) and axis ratio ($a/b$), and
the likelihoods of de Vaucouleurs and exponential fits to the galaxy
light profiles.  The non-detection of emission lines is determined
using the emission line measurements of \citet{yan06} by requiring
galaxies to have emission line fluxes below a 2$\sigma$ detection in
both the H$\alpha$ and the [O\textsc{ii}]$\lambda3727$ lines.  The
spectra used in this analysis are downloaded from the SDSS DR6 Data
Archive Server\footnote{http://das.sdss.org/DR6-cgi-bin/DAS}.

The galaxy radii are critical measurements for this analysis, entering
the FP determinations both directly through $R_e$ and indirectly
through the calculation of $I_e$.  We have attempted to correct for
known problems in the SDSS photometry pipeline (see Paper I) but
problems may remain.  To check the validity of the de Vaucouleurss
$R_e$ available in DR6, we compared these values to effective radii
measured independently using the photometry package GALFIT v.2.0.3c
\citep{peng02}, kindly provided by A. van der Wel.  The GALFIT radii
were measured from Sersic fits to the galaxy $r$-band light profile
with $n = 4$ to match the SDSS de Vaucouleurss photometry.  In
comparison to the GALFIT radii, the SDSS radii are slightly
underestimated for the largest galaxies (low by $\sim 0.07$ dex) and
slightly overestimated for the smallest galaxies (high by $\sim 0.1$
dex).  In addition to these systematic variations, there is scatter of
$\sim 0.05$ dex between the two estimates of $\log R_e$.  In
\S\ref{binning}, galaxies are sorted into bins based on $R_e$, with
bin widths of 0.2 dex in $\log R_e$.  Uncertainties on the order of
0.05 dex in $\log R_e$ should therefore have only modest effects on
bin assignments.  To make certain that our results are not strongly
affected by possible errors in the DR6 values of $R_e$, we have
performed the entire analysis presented in this paper using both the
SDSS pipeline values of $R_e$ and the GALFIT fits to $R_e$ and
compared the differences.  All the results are qualitatively
identical.  We have chosen to present the results based on the SDSS
DR6 pipeline photometry as these values are more easily accessible to
the general astronomical community.

\begin{figure}[t]
\includegraphics[width=1.0\linewidth]{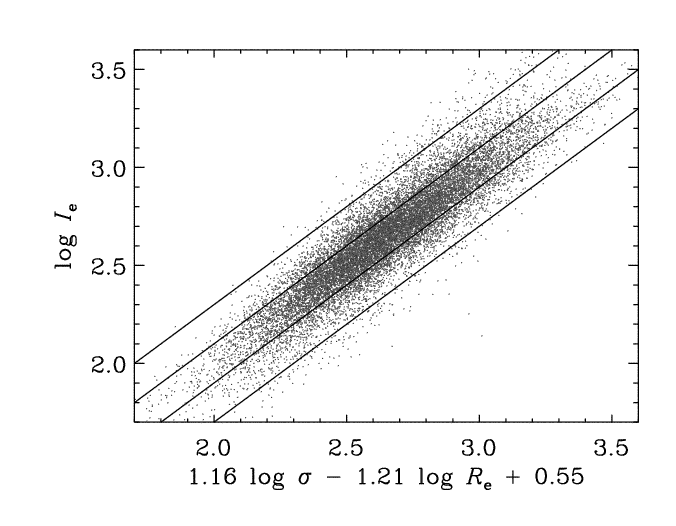}
\caption{The thickness of the FP in the $I_e$ dimension.  Here, the
  measured value of $\log I_e$ for each galaxy in the sample is
  plotted against the predicted $log I_e$ in terms of $\log \sigma$
  and $\log R_e$ from the best fit to the FP.  We divide the FP into
  ``low-SB'' ($-0.3 < \Delta \log I_e < -0.1$), ``midplane'' ($-0.1 <
  \Delta \log I_e < 0.1$), and ``high-SB'' ($0.1 < \Delta \log I_e <
  0.3$) slices based on the surface brightness residuals, as indicated
  by the solid lines.
  }\label{ibins}
\end{figure}

\section{Deconstructing the Fundamental Plane}\label{deconstructing_fp}

The overarching goal of this project is to understand the relation
between galaxy star formation histories (SFHs) and their present-day
structural properties, as categorized by their location in FP-space.
To study systematic variations in SFHs with structure, galaxies are
sorted into bins based on their location on the FP, then average
spectra are constructed from the individual galaxies in each bin.  The
average spectra have very high signal-to-noise ($S/N$), allowing
accurate stellar population modelling of the mean SFH of galaxies at
each point in FP-space.  This section defines the binning and
describes the stellar population modelling process.

\begin{figure*}[t]
\includegraphics[width=1.0\linewidth]{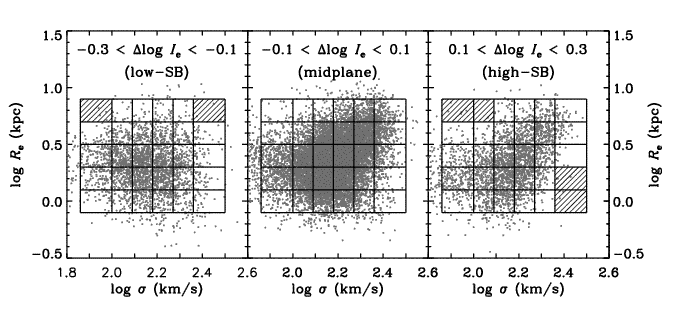}
\caption{Bin definitions for the purposes of sorting and stacking
  galaxies across the FP.  Gray points show all galaxies in the
  sample, while black lines indicate the bin definitions.  The three
  panels represent the three slices in surface brightness residuals,
  defined in Figure \ref{ibins}.  Galaxies are further divided into 6
  bins in $\log \sigma$ and 5 bins in $\log R_e$ within each slice in
  $\Delta \log I_e$, as illustrated.  Five of the 90 bins do not
  contain enough galaxies to produce robust mean spectra and are
  therefore excluded from our analysis.  These are indicated by the
  shaded bins.  In each of the three surface brightness slices, the
  galaxies occupy somewhat different parts of $\log \sigma$--$\log
  R_e$ space, indicating some curvature in the observed FP for this
  sample.  
}\label{islices}
\end{figure*}

\subsection{Binning in Fundamental Plane Space}\label{binning}

In addition to studying stellar population trends over the FP, we also
want to investigate variations through the thickness of the FP.  We
thus want to divide galaxies into bins in a three dimensional (3-D)
space.  There are two obvious choices for binning coordinate systems.
The first is to use two orthogonal vectors within the FP, together
with a third vector defined perpendicular to the FP, similar to the
$\kappa$-space of \citet{bender92}.  The second is to use the observed
structural parameters $\log \sigma$, $\log R_e$, and $\log I_e$
directly, despite the fact that they are not truly orthogonal.  We
have chosen the latter approach because the resulting trends yield
more readily to interpretation.

The parameters $\log \sigma$ and $\log R_e$ fortunately describe a
relatively face-on view of the FP.  It is furthermore attractive to
choose them as fundamental binning parameters because neither depends
explicitly on stellar population properties (in contrast to $I_e$) and
because a combination of $\sigma$ and $R_e$ gives an estimate of the
dynamical mass ($M_{dyn}$) of a galaxy, thus this binning scheme
groups together galaxies of similar total mass.  The FP by definition
has only two independent parameters.  Once $\log \sigma$ and $\log
R_e$ have been set, the mean value of $\log I_e$ for that bin is
fixed.

We use the $\log I_e$ dimension to explore SFH variations through the
{\it thickness} of the FP.  Treating $\log I_e$ as the dependent
variable, we fit the FP for $\log I_e$ as a function of $\log \sigma$
and $\log R_e$ using a simple least squares fit that minimizes
residuals in the $\log I_e$ direction (implemented as the IDL routine
{\it sfit.pro}) and find the relation
\begin{equation}\label{fp_relation}
\log I_e = 1.16 \log \sigma - 1.21 \log R_e + 0.55,
\end{equation}
where $\sigma$, $R_e$, and $I_e$ are in units of km s$^{-1}$, kpc, and
$L_{\odot}$ pc$^{-2}$, respectively.  The thickness of the FP can then
be described as $\Delta \log I_e$, where $\Delta$ indicates the
difference between the observed value of $\log I_e$ and the expected
value of $\log I_e$ given the observed $\log \sigma$ and $\log R_e$.
Figure \ref{ibins} shows the observed $\log I_e$ plotted against the
best fitting FP relation given by equation \ref{fp_relation}.  By
defining cuts in $\Delta \log I_e$, the FP is sliced into three
layers: ``low-SB'' ($-0.3 < \Delta \log I_e < -0.1$), ``midplane''
($-0.1 < \Delta \log I_e < +0.1$), and ``high-SB'' ($+0.1 < \Delta \log
I_e < +0.3$), as shown in Figure \ref{ibins}.  Here, the terms
``low-SB'' and ``high-SB'' are relative terms which describe the value
of $I_e$ with respect to the typical value of $I_e$ for galaxies of
the same $\sigma$ and $R_e$. 

More sophisticated statistical methods can be used to fit the FP
relation, but the assignment of galaxies to bins is relatively
insensitive to the exact equation used to describe the FP.  The choice
of $\log I_e$ as the dependent parameter in the binning process is
motivated by three things: the first is the previously stated fact
that $I_e$ (and not $R_e$ or $\sigma$) is expected to vary most
strongly with star formation history.  The second is that $\sigma$ and
$R_e$ can be combined to estimate $M_{dyn}$ for a galaxy, thus
producing groups of galaxies of similar total mass.  The third is that
Paper I demonstrated that, at fixed $\sigma$, stellar populations vary
substantially with $\Delta \log L_V$, which is related to $\Delta \log
I_e$.

In each of the three surface-brightness slices, galaxies are divided
by a $6 \times 5$ grid in $\log \sigma$ and $\log R_e$, as shown in
Figure \ref{islices}.  The grid is identical for all three slices.
Sorting galaxies in this 3-D space of $\log \sigma$--$\log
R_e$--$\Delta \log I_e$ results in 90 different bins of galaxies.
Within each bin, galaxies share similar values of these three
parameters, and therefore also share similar values of $M_{dyn}$,
total luminosity ($L_V$), and mass-to-light ratio ($M_{dyn}/L$).
Within each bin, galaxies with colors more than 0.06 mag\footnote{This
  is the typical color spread for galaxies with the same $\sigma$.
  See section \S4.1 of Paper I for details.} from the mean $g-r$ color
of the bin are excluded so that the stellar population properties
measured in each bin are not biased by a small number of interlopers.
The median values of $\log \sigma$, $\log R_e$, $\Delta \log I_e$, and
$\log I_e$ for each of the 90 bins are listed in Table 1,
along with the total number of galaxies in each bin.

Interestingly, Figure \ref{islices} shows that the galaxy distribution
in $\log \sigma$--$\log R_e$ space is somewhat different in the
various slices, with the high-SB slice showing a tail of
high--$\sigma$, high--$R_e$ galaxies that are not present in the
low-SB slice.  This occurs because the FP of our sample galaxies has a
small degree of curvature.  This curvature is not necessarily real and
may be due to selection effects.  We have not attempted to correct for
selection biases in this analysis, which spans a limited range in
redshift and is therefore nearly volume-limited.  Because galaxies are
binned by surface brightness, selection effects should result in
low-SB bins being incomplete but should not substantially bias the
mean spectra.

Of the 90 bins described here, six bins do not contain enough galaxies
to produce a reliable stacked spectrum (i.e., contain 5 or fewer
galaxies), leaving a total of 84 stacked spectra for the analysis.
Spectra are combined using a method that rejects outlier pixels, masks
bright skylines, weights all galaxies equally, and smoothes all
spectra to the same resolution before combining (see \S4.2 of Paper I
for details).  The result is a high $S/N$ mean spectrum for each bin
on the FP, as well as a corresponding error spectrum.

The bins are not evenly populated.  As can be seen in Figure
\ref{islices} and Table 1, some of the bins contain many
hundreds of galaxies, while others contain a dozen or so.  The typical
$S/N$ of the individual stacked spectra therefore varies substantially
between bins.  It is worth bearing in mind that the error bars are not
uniform for the stacked spectra.  In much of the subsequent analysis,
we will indicate data derived from the lower $S/N$ spectra and present
typical error bars separately for the low-$S/N$ and high-$S/N$ data.  

\subsection{Lick Indices and Stellar Population Modelling}\label{modelling}

In each of the stacked spectra, we measure the full set of Lick
indices as defined in \citet{worthey94} and \citet{worthey97}.  These
include Balmer lines H$\beta$, H$\gamma_F$, and H$\delta_F$ (as well
as broad versions of the bluer Balmer lines, H$\gamma_A$ and
H$\delta_A$), a set of Fe-dominated lines (Fe4383, Fe4531, Fe5015,
Fe5270, Fe5335, Fe5406, Fe5709, and Fe5782), and lines that are also
sensitive to element abundances other than Fe (Mg$_1$, Mg$_2$, Mg~$b$,
CN$_1$, CN$_2$, C$_2$4668, Ca4227, Ca4455, NaD, TiO$_1$, and TiO$_2$).
Index measurements are performed by the automated IDL code {\it
Lick\_EW}, which is part of the publicly
available\footnote{http://www.ucolick.org/$\sim$graves/EZ\_Ages.html}
{\it EZ\_Ages} code package \citep{graves08}.  {\it Lick\_EW} also
computes errors in the Lick index absorption strengths from the
associated error spectra following the formalism of \citet{cardiel98}.
It should be noted that the SDSS $3''$ spectral fibers sample roughly
$2/3$ $R_e$ for the redshift range of our sample.  Index measurements
and derived abundances in this work are therefore more nearly
comparable to literature values measured within $R_e$ or $R_e/2$,
rather than central or $R_e/8$ values.  Because many early-type
galaxies show strong radial metallicity gradients, central index
strengths and abundances tend to be higher than those measured from a
substantial fraction of the galaxy light.

All of the stacked spectra have been smoothed to match the
highest-$\sigma$ galaxies in the sample (300 km s$^{-1}$) in order to
compare all galaxies at the same effective spectral resolution.  The
combined smoothing due to $\sigma$ and the SDSS native resolution is
{\it lower} than the resolution at which the Lick indices are defined
($\sim 210$ km s$^{-1}$ at SDSS resolution); thus the linestrength
measurements must be corrected to bring them onto the Lick system.  We
do this using corrections computed in Table A2a of \citet{schiavon07}.
The stellar population models used to interpret the line strength
measurements are based on flux calibrated spectra, and the SDSS
spectra are themselves flux-calibrated; thus we do not apply a
zeropoint shift to bring our index measurements fully onto the Lick
system.  Because the data and the models are both flux-calibrated,
zeropoint offsets should be negligible.

\begin{figure}[t]
\includegraphics[width=1.0\linewidth]{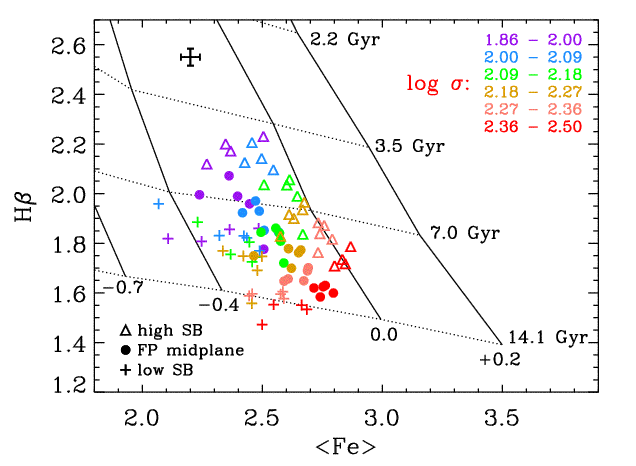}
\caption{Lick index measurements for the stacked spectra, compared
  with a grid of stellar population models.  The stellar population
  models shown here are computed using the average abundance pattern
  for the sample galaxies ([Mg/Fe] = $+0.19$, [C/Fe] = $+0.17$, [N/Fe]
  = $+0.05$, [Ca/Fe] = $+0.01$, with [Na/Fe] = [Si/Fe] = [Ti/Fe] =
  [Mg/Fe], and [Cr/Fe] = 0.0.  See \citealt{graves08} for details).
  Solid lines connect models at constant [Fe/H], while dotted lines
  connect models at constant age.  A typical statistical error for the
  index measurements is shown in the top left.  Lines of constant age
  (metallicity) do not run exactly horizontal (vertical) because the
  H$\beta$ and $\langle$Fe$\rangle$ indices are sensitive to both age
  and metallicity (the ``age-metallicity deneracy'').  As is clear
  from the model grids, this degeneracy can be broken using a
  combination of Lick indices.  The non-orthogonality of the model
  grids results in correlated errors in the age and [Fe/H]
  determinations for the stacked spectra, as discussed in
  \S\ref{residuals}.  }\label{grid_demo}
\end{figure}

\begin{figure*}[t]
\begin{center}
\includegraphics[width=0.75\linewidth]{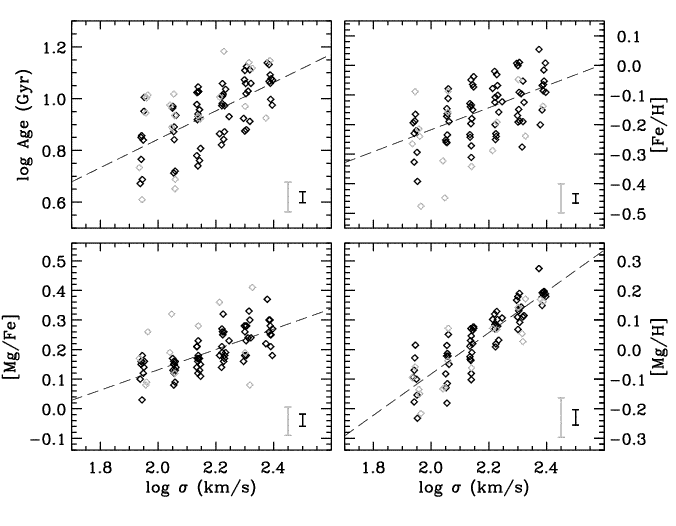}
\caption{Stellar population modelling results, showing mean
  luminosity-weighted stellar age, [Fe/H], [Mg/Fe], and [Mg/H] as a
  function of galaxy $\sigma$.  Black (gray) points and error bars indicate
  high (low) $S/N$ measurements and their associated median errors.
  Dashed lines show linear least squares fits of the stellar
  population properties as function of $\sigma$, based on the high
  $S/N$ (black) data only.  Mean stellar age, [Fe/H], [Mg/Fe], and
  [Mg/H] all increase with increasing $\sigma$.  Age and [Fe/H] both
  show substantial scatter at fixed $\sigma$, with total spread 4--5
  times the expected spread due to measurements errors, indicating
  genuine underlying population variations at fixed $\sigma$.  [Mg/H] and
  [Mg/Fe] show less scatter, only $\sim 2$ times that expected due to
  measurement errors.  The [Mg/H]--$\sigma$ relation is particularly
  strong and tight, nearly consistent with measurement errors,
  particularly at the high--$\sigma$ end.  
}\label{sig_v4}
\end{center}
\end{figure*}

To convert the set of line strength measurements into stellar
population mean ages and abundances, the Lick indices are compared to
the single stellar population models of \citet{schiavon07}.  These
models include the effects of multiple element abundance ratios and
allow estimates of [Fe/H], [Mg/Fe], [C/Fe], [N/Fe], and [Ca/Fe], as
well as mean luminosity-weighted stellar age.  In \citet{graves08}, we
created an IDL code package, {\it EZ\_Ages}, which automates the
process of determining stellar population mean age, [Fe/H], and
abundance ratios for a set of Lick index measurements.  We use {\it
EZ\_Ages} with the ``standard set'' of Lick indices (H$\beta$,
$\langle$Fe$\rangle$\footnote{$\langle$Fe$\rangle$ = $\frac{1}{2}$
(Fe5270 + Fe5335)}, Mg~$b$, C$_2$4668, CN$_1$, and Ca4227) to
determine ages, [Fe/H], and abundance ratios for the 84 stacked
spectra in our sample with adequate $S/N$ (see
\S\ref{binning})\footnote{For stacked spectrum \#76, the H$\beta$ line
was weaker than any of the models.  We therefore computed abundances
for this spectrum using H$\gamma_F$ instead of H$\beta$.  Because
H$\gamma_F$ produces systematically younger age estimates than
H$\beta$ \citep{graves08}, we have excluded this spectrum from the age
analysis while retaining it for the abundance analysis.}.  A detailed
description of the modelling process is presented in \citet{graves08},
as well as rigorous testing of the method on Galactic globular
clusters and a comparison with stellar population modelling results
from \citet{thomas05}.  The interested reader is referred to that work
for more information on the modelling process.  It is worth noting
here, however, that the models are computed at fixed [Fe/H], rather
than at fixed total metallicity ([Z/H]) as in some stellar population
models.  Total metallicity is dominated by oxygen, which is
unmeasurable in unresolved stellar populations with current
techniques.  In this analysis, we specifically discuss [Fe/H] and
[Mg/H], elements which can be measured from low-resolution spectra,
and avoid making statements about [Z/H].

Figure \ref{grid_demo} shows the measured values of
$\langle$Fe$\rangle$ and H$\beta$ for the stacked spectra plotted over
a set of model grids from \citet{schiavon07}.  Data points are
color-coded by $\sigma$ as indicated in the figure key.  Crosses,
filled circles, and open triangles represent the low-SB, midplane, and
high-SB data, respectively.  The various bins in $R_e$ are not
indicated.  Two trends emerge clearly from the data.  As $\sigma$
increases, galaxies tend to have {\it stronger} $\langle$Fe$\rangle$
and {\it weaker} H$\beta$.  At fixed $\sigma$, as $\Delta I_e$
increases, galaxies tend to have {\it stronger} $\langle$Fe$\rangle$
and {\it stronger} H$\beta$.  We will discuss the detailed
implications of these trends in \S\ref{stellar_pops}; here they are
merely presented to illustrate the linestrength data.

The solid lines in Figure \ref{grid_demo} show models of constant
[Fe/H], from $-0.7$ to $+0.2$, as labelled.  The dotted lines show
models of constant age, from 2.2 to 14.1 Gyr.  As can be seen from the
orientation of the grid lines, H$\beta$ is predominantly sensitive to
stellar population age such that younger populations have stronger
H$\beta$, while $\langle$Fe$\rangle$ is dominated by [Fe/H] such that
more Fe-rich populations have stronger $\langle$Fe$\rangle$.  The
effects of the age-metallicity degeneracy are apparent in the fact
that H$\beta$ is (mildly) sensitive to [Fe/H] and $\langle$Fe$\rangle$
is sensitive to age.  Because of the degeneracy, there is no
one-to-one mapping between H$\beta$ and age, or $\langle$Fe$\rangle$
and [Fe/H]; the grid lines are not perfectly vertical and horizontal.
However, the model grids in these particular indices are close enough
to orthogonal that they serve to break the age-metallicity degeneracy.
Other combinations of indices (e.g., H$\delta_F$ and Fe4383, see
Figure 9 of \citealt{graves08}) are more affected by the degeneracy,
causing the model grids to collapse down on top of one another and
making them less robust indicators of age and metallicity.  

\begin{figure*}[t]
\begin{center}
\includegraphics[width=0.75\linewidth]{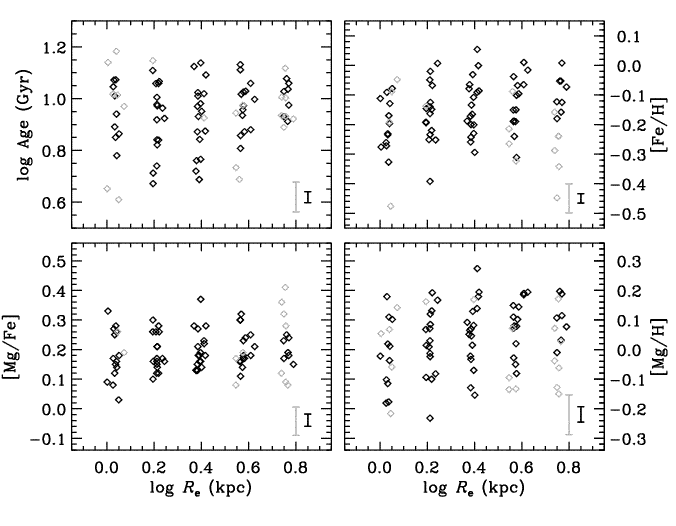}
\caption{Stellar population modelling results, showing mean
  luminosity-weighted stellar age, [Fe/H], [Mg/Fe], and [Mg/H] as a
  function of galaxy $R_e$.  Black (gray) points and error bars
  indicate high (low) $S/N$ measurements and their associated median
  errors.  No strong correlations are observed between $R_e$ and any
  stellar population properties, implying that galaxy star formation
  histories are nearly independent of galaxy size.  }\label{reff_v4}
\end{center}
\end{figure*}

{\it EZ\_Ages} computes statistical errors in each of the stellar
population parameters, determined from the measurement errors in the
Lick indices.  The black cross in the upper left corner of Figure
\ref{grid_demo} illustrates the median errorbars in the line strenth
measurements.  The non-orthogonality of the model grids due to the
age-metallicity degeneracy causes the derived errors to be correlated.
For example, an overestimate of $\langle$Fe$\rangle$ will result in
both a higher derived value of [Fe/H] and a slightly lower derived
value for the age.  The age-metallicity degeneracy therefore causes
correlated errors in the stellar population parameters, which must be
taken into account when interpreting results.  In general, however,
the very high $S/N$ of the stacked spectra in this analysis make the
stellar population modelling results robust to such effects.  The
effect of correlated errors will be discussed in greater detail in
\S\ref{residuals}.

This paper discusses only the mean luminosity-weighted ages, [Fe/H], the
abundance ratio [Mg/Fe], and [Mg/H] (which is simply [Mg/Fe] +
[Fe/H]).  The results of the stellar population fits are given in
Table 2, along with the Lick indices from which these
parameters are determined.  Discussion of other abundance ratios is
deferred to future work.  An important caveat about the stellar
population parameters discussed here is that they represent mean,
luminosity-weighted averages over all the stars in a galaxy, which
additionally have been averaged over all galaxies which contribute to
a stacked spectrum.  A younger mean age measured in one stacked
spectrum compared to another does not make a definitive statement
about the ages of all stars in all galaxies in that bin: it could also
mean that all of the galaxies in that bin contain sub-populations of
younger stars, or that some of the galaxies in the bin have generally
younger stars.  There is also a strong degeneracy between the age of a
younger subpopulation and the mass fraction it comprises: mean ages
can be skewed younger by a small population of very young stars, or by
a larger population of somewhat less young stars.  Age measurements
should therefore be treated as a statistical description of galaxies
in the bin, rather than as ``true'' ages for all the stars in an
individual galaxy.

\section{Stellar Populations in Fundamental Plane Space}\label{stellar_pops}

\subsection{Age, [Fe/H], [Mg/H], and [Mg/Fe] as Functions of
  $\sigma$, $R_e$, and $I_e$ Separately}\label{stellar_pops_v3}

We begin by examining the variations in stellar population properties
with each of the FP parameters $\sigma$, $R_e$, and $I_e$
individually.  

Figure \ref{sig_v4} shows the {\it EZ\_Ages} results for mean
luminosity-weighted age, [Fe/H], [Mg/H], and [Mg/Fe] as a function of the
median $\log \sigma$ in each galaxy bin.  We have divided the data
into ``low-$S/N$'' and ``high-$S/N$'' subsamples in order to indicate
the range of error values, where low-$S/N$ data points are those whose
statistical error estimates are $> 0.015$ dex higher than the median
error value for the data as a whole.  The black data points and
error bars indicate the values associated with high-$S/N$ data.
Low-$S/N$ data are indicated in gray, along with the appropriate
median error bar.  Age, total Fe and Mg abundances, and the [Mg/Fe]
abundance ratio all increase with increasing $\sigma$.  These
relations are very similar to those demonstrated in Paper I, where
galaxies were binned first by $\sigma$ and then by $L$ and color at
fixed $\sigma$.  Dashed lines indicated linear least squares fits of
the stellar population parameters as functions of $\sigma$, computed
using only the high $S/N$ data (i.e., excluding the gray points).

\begin{figure*}[t]
\begin{center}
\includegraphics[width=0.75\linewidth]{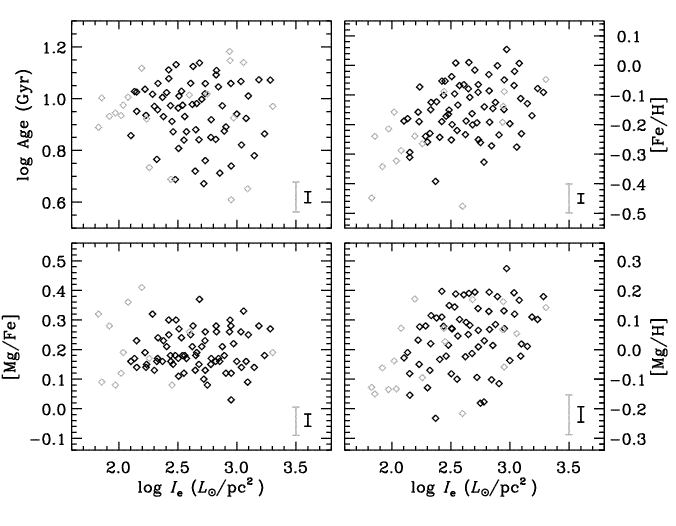}
\caption{Stellar population modelling results, showing mean
  luminosity-weighted stellar age, [Fe/H], [Mg/Fe], and [Mg/H] as a
  function of galaxy $I_e$.  Black (gray) points and error bars indicate
  high (low) $S/N$ measurements and their associated median errors.
  Weak trends are visible, such that stellar population age decreases
  slightly with increasing $I_e$, while [Fe/H] and [Mg/H] increase
  slightly with increasing $I_e$, but none of these trends are
  compelling.  [Mg/Fe] appears entirely independent of $I_e$.  
}\label{ieff_v4}
\end{center}
\end{figure*}

The age--$\sigma$ and [Fe/H]--$\sigma$ relations both show substantial
scatter.  Quantitatively, the spread around the mean relations (dashed
lines) is $\sim$4--5 times larger than the scatter expected due to the
statistical errors in both age and [Fe/H], indicating that there is
substantial spread in the star formation histories of galaxies at
fixed $\sigma$.  The [Mg/Fe]--$\sigma$ and [Mg/H]--$\sigma$ relations
show less scatter, with the observed spread representing only $\sim$2
times the expected statistical scatter.  The Mg abundance shows both
the steepest correlation with $\sigma$ and the least intrinsic scatter
about the mean relation as compared to the observational errors,
indicating that [Mg/H] varies only slightly at fixed $\sigma$.

Unlike the strong stellar population trends observed with $\sigma$,
neither $R_e$ nor $I_e$ shows strong correlations with stellar
population properties.  Figure \ref{reff_v4} shows the results of the
stellar population modelling as a function of the median $\log R_e$
for each galaxy bin.  There are no strong trends with $R_e$ in any of
the properties modelled here.  There is some evidence for slight
increases in [Fe/H] and [Mg/H] with increasing $R_e$, but the
dependence is weak and the scatter is large.  Also, at large $R_e$
there do not appear to be any ``young'' galaxies with mean ages $< 7$
Gyr.  However, it is evident from Figure \ref{islices} that there are
very few galaxies with large $R_e$ and low $\sigma$.  Since Figure
\ref{sig_v4} showed that low--$\sigma$ galaxies are most likely to
have young ages, the lack of ``young'' galaxies with large $R_e$ may
be due to the way galaxies populate the FP, rather than any genuine
relation between larger $R_e$ and older ages.  

Finally, Figure \ref{ieff_v4} shows age, [Fe/H], [Mg/Fe], and [Mg/H]
plotted against the median $\log I_e$ for each galaxy bin.  Again,
there are no strong correlations between $\log I_e$ and any of the
stellar population parameters.  There is some indication of weak
correlations between $\log I_e$ and [Fe/H] and [Mg/H] such that the
high-SB galaxies have higher Fe and Mg abundances.  There is also some
evidence for a weak anti-correlation between $\log I_e$ and age such
that low-SB galaxies have older ages, but none of these trends is
compelling.  The [Mg/Fe] abundance ratio is completely flat with $\log
I_e$.

A clear result from Figures \ref{sig_v4}--\ref{ieff_v4} is that
stellar populations on the FP vary primarily with $\sigma$.  Neither
$R_e$ nor $I_e$ appears to be closely related to stellar population
properties.  However, the lack of general trends with $R_e$ and $I_e$
may hide weak correlations with residuals from the strong $\sigma$
relationships.  Although Figure \ref{sig_v4} shows clear correlations
with $\sigma$ in all stellar population parameters, there is some
scatter at fixed $\sigma$, particularly in age and [Fe/H].  The next
section examines how stellar populations vary throughout the FP as a
function of all three FP parameters simultaneously.

\begin{figure*}[t]
\begin{center}
\includegraphics[width=1.0\linewidth]{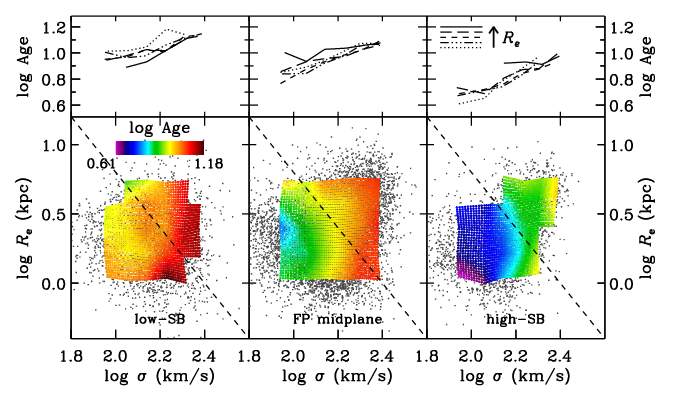}
\caption{Mean luminosity-weighted stellar population age, mapped
  across the FP.  The three lower panels show the three slices in
  surface brightness, as in Figure \ref{islices}.  Within each slice,
  galaxy $\sigma$ and $R_e$ are plotted, with gray points showing
  individual galaxies in the sample.  The overplotted color contours
  show the typical stellar population age at each point on the FP (see
  \S\ref{mapping} for details).  The dashed lines show lines of
  constant $M_{dyn}$, assuming $M_{dyn} \propto \sigma^2 R_e$.  Within
  each slice, age increases with increasing $\sigma$, as seen in
  Figure \ref{reff_v4}.  Lines of constant age run approximately
  vertically, indicating that mean stellar population age is
  independent of $R_e$ at fixed $\sigma$.  The upper panels show
  stellar population age as a function of $\sigma$ for each FP slice.
  Different values of $R_e$ are indicated by different line styles.
  Comparing the typical ages between the various slices in $I_e$, it
  is clear that galaxies with lower surface brightness than the FP
  (left panels) have {\it older} ages than those on the FP midplane
  (center panels), while galaxies with higher surface brightness than
  the FP (right panels) have {\it younger} ages, at the same value of
  $\sigma$ and $R_e$.  Despite the lack of strong stellar population
  trends with absolute galaxy $I_e$ (c.f., Figure \ref{ieff_v4}),
  stellar population age at fixed $\sigma$ does appear to depend on
  $I_e$ residuals from the FP, such that low-SB (high-SB) galaxies are
  older (younger) than those on the FP midplane.  }\label{age_map}
\end{center}
\end{figure*}

\subsection{Mapping Stellar Populations Along and Through the Fundamental
  Plane}\label{mapping}

By binning galaxies along and through the thickness of the FP, then
stacking their spectra and performing stellar population modelling on
the high $S/N$ average spectra, we have measured the typical age,
[Fe/H], [Mg/H], and [Mg/Fe] for galaxies at each point on, above, and
below the FP.  Each of these properties can be mapped throughout the
3-D space of the FP, thereby associating information about typical
star formation histories with the structural properties of galaxies.
This is a somewhat challenging exercise in data visualization, as we
are trying to understand the variations and co-variations of four
different stellar population parameters throughout a 3-D structure
space.  This effectively involves probing a 7-D parameter space,
although by focusing on one stellar population parameter at a time, we
can limit it to a 4-D space.  In this and the next section, we attempt
to codify trends in this 7-D space and understand how the various
galaxy properties correlate with one another.  We begin by mapping
each of the four stellar population properties along and through the
FP.

\begin{figure*}[t]
\begin{center}
\includegraphics[width=1.0\linewidth]{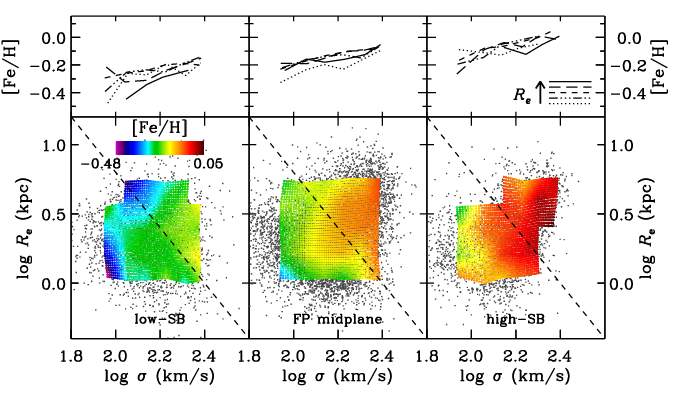}
\caption{Mean stellar population [Fe/H], mapped across the FP.  The
  three lower panels show the three slices in surface brightness, as
  in Figure \ref{islices}.  Within each slice, galaxy $\sigma$ and
  $R_e$ are plotted, with gray points showing individual galaxies in
  the sample.  The overplotted color contours show the typical [Fe/H]
  at each point on the FP.  The dashed lines show lines of constant
  $M_{dyn}$, assuming $M_{dyn} \propto \sigma^2 R_e$.  Within each
  slice, [Fe/H] increases with increasing $\sigma$, as in Figure
  \ref{sig_v4}.  [Fe/H] appears to be relatively independent of $R_e$,
  with contours of equal [Fe/H] running nearly vertically in all
  slices.  The upper panels show [Fe/H] as a function of $\sigma$ for
  each FP slice.  Different values of $R_e$ are indicated by different
  line styles.  Comparing the various slices in surface brightness,
  low-SB galaxies (left panels) appear to have {\it lower} [Fe/H] than
  galaxies on the midplane of the FP (center panels) at the same
  $\sigma$ and $R_e$, while high-SB galaxies (right panels) have
  higher [Fe/H].  Combining these results with those apparent in
  Figure \ref{age_map} suggests that, at fixed $\sigma$, stellar
  population age and [Fe/H] are {\it anti-correlated}.
  }\label{feh_map}
\end{center}
\end{figure*}

Figure \ref{age_map} shows the results for mean luminosity-weighed
age.  The figure is constructed to be similar to Figure \ref{islices},
which illustrated the bin definitions for the 3-D FP space.  The three
lower panels show the three slices in $\Delta \log I_e$, with the
center panel showing the midplane slice through the FP, while the left
and right panels show respectively the low-SB and high-SB slices from
Figure \ref{ibins}.  Within each FP slice, $\log \sigma$ and $\log
R_e$ are plotted for the individual galaxies as gray points (as in
Figure \ref{islices}).  Dashed lines show lines of constant $M_{dyn}$.
The color overlay indicates the mean stellar population age at each
point in the parameter space, with numerical values given by the color
bar in the top of the leftmost panel.  The age overlay was constructed
as follows: for each stacked spectrum, the median values of $\log
\sigma$ and $\log R_e$ are plotted for the galaxies in the bin,
color-coded by the age measured in the stacked spectrum.  Ages are
then interpolated between the median points to create a continuous map
of stellar population age across the $\log \sigma$--$\log R_e$
diagram.  High-$S/N$ and low-$S/N$ data are all included in the
interpolation.

The upper three panels of Figure \ref{age_map} show mean
luminosity-weighted age as a function of $\sigma$ for each FP slice.
Different bins in $R_e$ are indicated by different line styles.

Look first at the central panels of Figure \ref{age_map}, which show
the midplane of the FP.  The trend from Figure \ref{sig_v4} is
visible: galaxies with larger $\sigma$ are older (red) than galaxies
with lower $\sigma$ (blue).  Although there are significant changes in
age as a function of $\sigma$, there seems to be little if any
variation in age as a function of $R_e$.  Lines of constant age run
roughly vertically in the diagram.  This is consistent with Figure
\ref{reff_v4}, which showed no clear systematic variation in stellar
population age as a function of $R_e$, independently of the other FP
parameters.  Here we see directly that, at fixed $\sigma$, there is no
substantial dependence of stellar population age on $R_e$.  If the
best predictor of stellar population age were $M_{dyn}$ rather than
$\sigma$, lines of constant age should follow the dashed line of
constant $M_{dyn}$.  The near-verticality of the age contours
indicates that $\sigma$ is a better predictor of stellar population
age than is $M_{dyn}$.

\begin{figure*}[t]
\begin{center}
\includegraphics[width=1.0\linewidth]{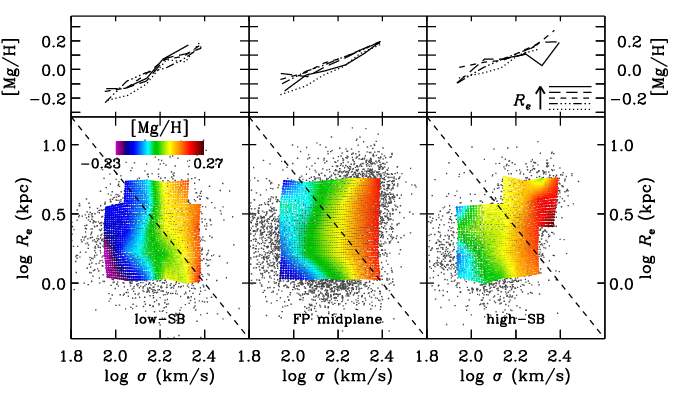}
\caption{Mean stellar population [Mg/H], mapped across the FP.  The
  three lower panels show the three slices in surface brightness, as
  in Figure \ref{islices}.  Within each slice, galaxy $\sigma$ and
  $R_e$ are plotted, with gray points showing individual galaxies in
  the sample.  The overplotted color contours show the typical [Mg/H]
  at each point on the FP.  The dashed lines show lines of constant
  $M_{dyn}$, assuming $M_{dyn} \propto \sigma^2 R_e$.  Within each
  slice, [Mg/H] increases strongly with increasing $\sigma$, as in
  Figure \ref{sig_v4}.  Like age and [Fe/H], [Mg/H] appears to be
  relatively independent of $R_e$, with contours of equal [Mg/H]
  running nearly vertically in all slices.  The upper panels show
  [Mg/H] as a function of $\sigma$ for each FP slice.  Different
  values of $R_e$ are indicated by different line styles.  Comparing
  the various slices of the FP, there appears to be little variation
  in [Mg/H] between the low-SB galaxies (left panels), the FP midplane
  galaxies (center panels), and the high-SB galaxies (right panels) at
  fixed $\sigma$ and $R_e$.  There is a slight tendency for high-SB
  galaxies to have higher [Mg/H], but the trend with $\Delta \log I_e$
  is substantially weaker than that observed for [Fe/H] in Figure
  \ref{feh_map}.  }\label{mgh_map}
\end{center}
\end{figure*}

If we look now at the galaxies from the low-SB (left panels) and
high-SB (right panels) slices above and below the FP, we see
essentially the same trends: age increases systematically with
increasing $\sigma$, but lines of constant age run nearly vertically,
indicating that stellar population age is independent of $R_e$ at
fixed $\sigma$.  However, comparing the age {\it ranges} (indicated by
the color scale) between the different panels, there are systematic
differences.  These are illustrated clearly in the upper panels.
Galaxies in the low-SB slice of the FP (left panels) are
systematically older than galaxies in the midplane of the FP (center
panels), which are in turn systematically older than galaxies in the
high-SB slice (right panels).  There are substantial overlaps in the
age ranges in the three panels, but {\it at fixed} $\sigma$ the low-SB
galaxies are older and the high-SB galaxies are younger than the
typical galaxies in the FP.  This means that, although there are no
clear trends in stellar population age with $I_e$ alone (Figure
\ref{ieff_v4}), typical galaxy ages do vary with $\Delta \log I_e$ at
a fixed point in $\sigma$ and $R_e$.  This is similar to the results
of \citet{forbes98} and \citet{terlevich02}, who found that surface
brightness residuals from the FP correlate with stellar population
age, although they explored only the general trend over all galaxies,
not as a function of $\sigma$ and $R_e$.  Our higher-resolution map is
made possible through the large sample of 16,000 galaxies in this
study.

A similar map for [Fe/H] is shown in Figure \ref{feh_map}.  As seen in
Figure \ref{sig_v4}, [Fe/H] increases systematically with $\sigma$.
Like stellar population age, [Fe/H] appears to depend weakly or
negligibly on $R_e$---contours of constant [Fe/H] are roughly vertical
rather than following the lines of constant $M_{dyn}$.  Also similar
to the age variations, there appear to be systematic variations in
[Fe/H] between the different slices in $\Delta \log I_e$.  This trend,
however, runs opposite to the one seen in age: at fixed $\sigma$, the
low-SB galaxies (left panels) tend to be Fe-poor and the high-SB
galaxies (right panels) tend to be Fe-rich.  The general conclusion is
that, not only do age and [Fe/H] increase with $\sigma$, they also
vary systematically with $\Delta \log I_e$, with low-SB galaxies
showing older mean ages and lower [Fe/H], while high-SB galaxies show
younger mean ages and higher [Fe/H].  This implies that, at fixed
$\sigma$, age and [Fe/H] are {\it anti}-correlated, as first shown by
\citet{trager00b}.  Section \ref{residuals} will directly examine the
correlations between various stellar population residuals from the
mean trends with $\sigma$.

Figure \ref{mgh_map} shows the map for [Mg/H].  As expected from
Figure \ref{sig_v4}, [Mg/H] varies strongly with $\sigma$.  Like
stellar population age and [Fe/H], [Mg/H] shows no dependence on
$R_e$.  Unlike age and [Fe/H], [Mg/H] varies only mildly with $\Delta
\log I_e$.  To the extent that [Mg/H] does vary at fixed $\sigma$
between the different panels, it follows the same trend as [Fe/H],
with low-SB galaxies showing slightly lower [Mg/H] and high-SB
galaxies showing slightly higher [Mg/H] than their counterparts in the
midplane FP slice.  However, the variation in [Mg/H] with $\Delta \log
I_e$ is only about half than that seen in [Fe/H].  Figure \ref{sig_v4}
showed that [Mg/H] has the strongest and tightest trend with $\sigma$.
This 4-D map both confirms that there is little spread in [Mg/H] at
fixed $\sigma$ and illustrates that the small existing spread is
correlated with $\Delta \log I_e$.

\begin{figure*}[t]
\begin{center}
\includegraphics[width=1.0\linewidth]{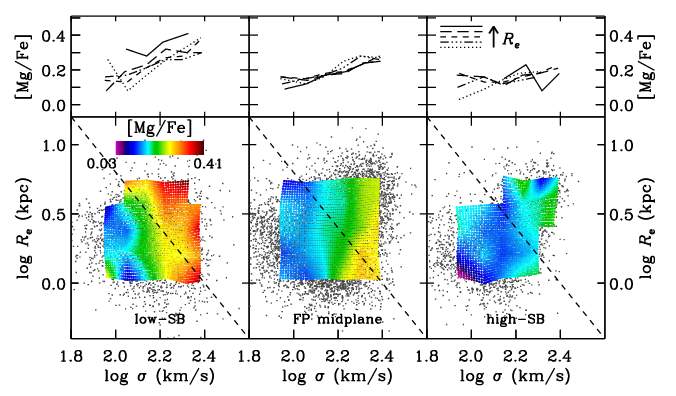}
\caption{Mean stellar population [Mg/Fe], mapped across the FP.  The
  three lower panels show the three slices in surface brightness, as
  in Figure \ref{islices}.  Within each slice, galaxy $\sigma$ and
  $R_e$ are plotted, with gray points showing individual galaxies in
  the sample.  The overplotted color contours show the typical [Mg/Fe]
  at each point on the FP.  The dashed lines show lines of constant
  $M_{dyn}$, assuming $M_{dyn} \propto \sigma^2 R_e$.  Within each
  slice, [Mg/Fe] increases strongly with increasing $\sigma$, as in
  Figure \ref{sig_v4}.  Like the other stellar population parameters,
  [Mg/Fe] appears to be relatively independent of $R_e$, with contours
  of equal [Mg/Fe] running nearly vertically in all slices.  The upper
  panels show [Mg/Fe] as a function of $\sigma$ for each FP slice.
  Different values of $R_e$ are indicated by different line styles.
  On average, low-SB galaxies (left panels) have higher [Mg/Fe] than
  those on the FP midplane (center panels), which in turn have higher
  [Mg/Fe] than the high-SB galaxies (right panels).  The low-SB
  galaxies show a strong increase in [Mg/Fe] with $\sigma$, while the
  high-SB galaxies show very little overall variation in [Mg/Fe].
  Another way to state this trend is that, at low $\sigma$ (leftmost
  galaxies in each panel), all galaxies have relatively low values of
  [Mg/Fe], while at high $\sigma$ (rightmost galaxies in each panel),
  there is a substantial range in [Mg/Fe].  }\label{mgfe_map}
\end{center}
\end{figure*}

Finally, Figure \ref{mgfe_map} shows the map of [Mg/Fe] across the FP
slices.  As with the other stellar population parameters, [Mg/Fe]
depends strongly on $\sigma$ and is independent of $R_e$.  Comparison
between the panels shows that it also varies with $\Delta \log I_e$.
Unlike [Fe/H] and [Mg/H], both of which {\it increase} with increasing
surface brightness, [Mg/Fe] {\it decreases} on average in high-SB
galaxies (right panels) and {\it increases} on average in low-SB
galaxies (left panels).  In this respect, [Mg/Fe] behaves similarly to
stellar population age, showing the highest values in low-SB, older
galaxies.

It is interesting that the low-SB slice of the FP (left panels) shows
substantially more total variation in [Mg/Fe] than do the other
panels.  This trend is consistent through all three slices.  Indeed,
the high-$I_e$ slice (right panels) shows only a small range of
[Mg/Fe], while the low-$I_e$ slice (left panels) shows substantial
variation.  An equally valid way to state this (although perhaps
harder to see clearly in Figure \ref{mgfe_map}) would be to say that
high--$\sigma$ galaxies (at the right side of each map) show
substantial variation in [Mg/Fe] between galaxies, but that the
low--$\sigma$ galaxies (at the left side of each map) all have
similarly low values of [Mg/Fe].

Taken together, these maps of stellar population parameters along and
through the FP show that age, [Fe/H], [Mg/H], and [Mg/Fe] all vary
strongly with $\sigma$, but that they also appear to vary with $\Delta
\log I_e$ ([Mg/H] varies only mildly, while the other three parameters
vary substantially).  None of them varies significantly with $R_e$.
From these maps, it is clear that the spread around the $\sigma$
relations shown in Figure \ref{sig_v4} is not just due to statistical
errors but in fact is systematic behavior that is correlated with
surface brightness differences.  This furthermore implies that there
are substantial correlations between stellar population properties at
fixed $\sigma$.  The next section looks explicitly at the correlations
between residuals from the $\sigma$-driven relations of Figure
\ref{sig_v4} in order to understand the systematic co-variation of the
various stellar population properties.

\begin{figure*}[t]
\begin{center}
\includegraphics[width=0.75\linewidth]{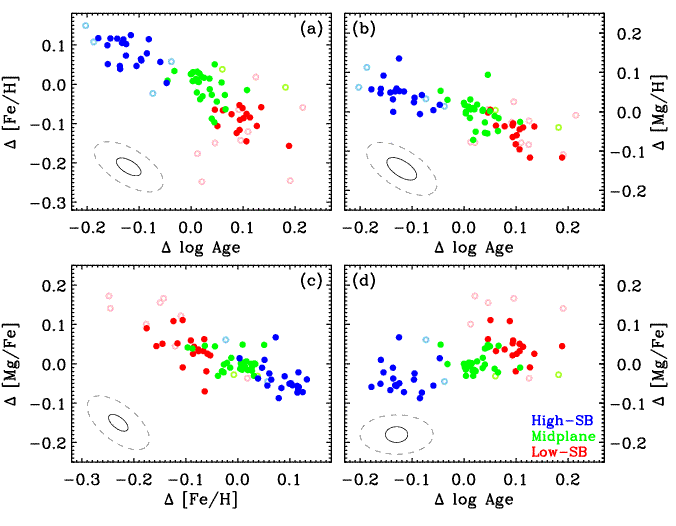}
\caption{Residuals from the mean stellar population trends with
  $\sigma$, color-coded by surface brightness residuals from the FP
  ($\Delta \log I_e$).  The $\Delta$ value for each stellar population
  parameter is determined by subtracting from the measured value the
  mean relation given by the dashed line in Figure \ref{sig_v4}.  In
  each panel, solid (open) circles and black (gray) 1-$\sigma$ error
  ellipses represent high (low) $S/N$ data and the corresponding
  measurement errors.  Error ellipses show the direction of the
  correlated errors in the age and abundance determinations.  Data are
  color-coded by galaxy surface brightness, with galaxies in the
  low-SB, midplane, and high-SB FP slices shown in red, green, and blue,
  respectively.  Stellar population residuals from the trends with
  $\sigma$ are clearly correlated with one another and with $\Delta
  \log I_e$ such that, at fixed $\sigma$, low-SB galaxies tend to be
  older, Fe-poor, Mg-poor, and enhanced in [Mg/Fe] compared to
  galaxies in the FP midplane, while high-SB galaxies are younger,
  Fe-rich, Mg-rich, and with lower [Mg/Fe].  }\label{resids}
\end{center}
\end{figure*}

\subsection{Residuals from the Stellar Population--$\sigma$ Relations}\label{residuals} 

Stellar population residuals $\Delta (\log$ age), $\Delta$[Fe/H],
$\Delta$[Mg/H], and $\Delta$[Mg/Fe] are defined with respect to the
mean relations versus $\sigma$ by subtracting the linear fits shown in
Figure \ref{sig_v4} (dashed lines).  Figure \ref{resids} shows various
correlations between these residuals.  Filled circles show the
high-$S/N$ data (corresponding to black points in Figure
\ref{sig_v4}), while empty circles show the low-$S/N$ data (gray
points in Figure \ref{sig_v4}).  Data points are color-coded by
$\Delta \log I_e$, with blue representing the high-SB slice of the FP,
green representing the midplane of the FP, and red showing the low-SB
slice of the FP.

As discussed in \S\ref{modelling}, the age-metallicity degeneracy
results in correlated errors in the stellar population modelling
process.  Correlated errors arise because all absorption indices are
sensitive to both age and metallicity, resulting in model grid lines
that are not orthogonal in index-index space (see Figure
\ref{grid_demo}).  The error ellipses in the lower left corners of
Figure \ref{resids} indicate the slope of the correlated errors from
the stellar population modelling, as determined by Monte Carlo
simulations in \citet[see Figure 3 of that work]{graves08}.  The solid
black and dashed gray error ellipses correspond to the median errors
for high- and low-$S/N$ data, respectively, based on the error
estimates for the stellar population parameters determined by {\it
EZ\_Ages}.

From Figure \ref{resids}, it is clear that there are significant
correlations between the residuals from the mean relations with
$\sigma$.  These correlations are strongly associated with $\Delta
\log I_e$ and thus represent real variations in star formation
histories for galaxies with the same $\sigma$.  Although the
correlated residuals in panels a--c are in the same direction as the
correlated errors from the stellar population modelling process, the
residuals trends cannot be explained merely by the correlated errors,
for two reasons.  In all cases, the black error ellipses are far too
small to account for the observed spread in residuals, so the
degeneracies in the modelling process cannot produce the observed
spread in the data.  Even more conclusively, the residuals depend
strongly on $\Delta \log I_e$.  This cannot be explained by
measurement errors in the absorption line strengths propagated through
the correlated errors in the stellar population modelling because the
sense of the measurement errors should be random, not dependent on
$\Delta \log I_e$.  Thus the known degeneracies of the modelling
process {\it cannot} account for the observed correlation of
residuals.  The appendix further demonstrates that correlated
systematic errors due to multi-burst stellar population models also
cannot be responsible for the observed trends.

Figure \ref{resids}a indicates that there is a strong {\it
  anti}-correlation between stellar population age and [Fe/H] at fixed
  $\sigma$, such that high-SB galaxies tend to be younger and more
  Fe-rich than their low-SB counterparts at the same $\sigma$.  The
  total spread is at least a factor of two (0.3 dex) in both age and
  Fe abundance, with $\Delta$[Fe/H] $\propto -0.7 \Delta (\log
  \mbox{age})$.  A similar anti-correlation was noted by
  \citet{trager00b} and more recently by \citet{smith07-iaus245},
  whose error estimates for individual galaxies were small enough to
  allow them to claim a genuine anti-correlation in the parameters.
  The clear dependence of age and [Fe/H] residuals on $\Delta \log
  I_e$ shown in Figure \ref{resids}a lays to rest any doubts that the
  anti-correlation could be due to errors alone, even though the slope
  of the anti-correlation is similar to the expected correlation of
  observational errors.  In contrast to the clear dependence on
  $\Delta \log I_e$ shown here, \citet{trager00b} did not find any
  dependence of the age--[Fe/H] anti-correlation on $I_e$ but in fact
  found that there was no improvement in statistical fits of age or
  [Fe/H] when $I_e$ was included with $\sigma$ as one of the fitting
  parameters.  This is probably because they were fitting against
  absolute $I_e$ rather than $\Delta \log I_e$ at fixed $\sigma$ and
  $R_e$ (recall that Figure \ref{ieff_v4} shows only very weak
  dependence of stellar population parameters on absolute $\log I_e$).
  There is a message here for multi-parameter fits in general: to
  detect a relation versus a residual in a higher-dimensional space,
  it is necessary to fit against that residual explicitly; the
  variation may remain concealed when using the parent coordinate
  alone.

\begin{figure}[t]
\includegraphics[width=1.0\linewidth]{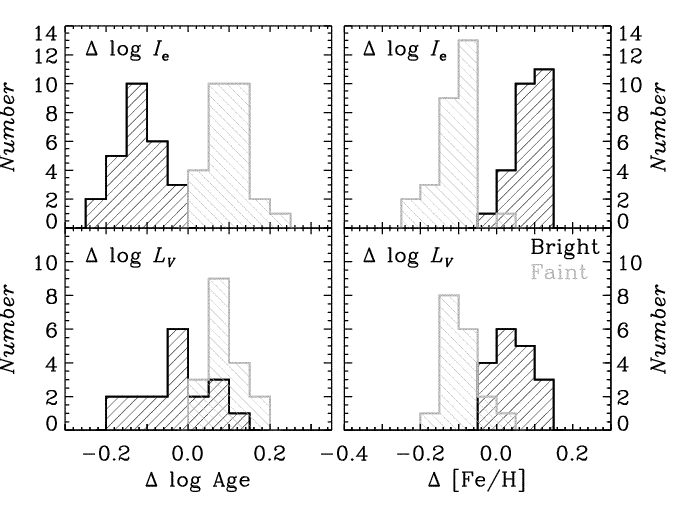}
\caption{Comparison of the discriminatory power of $\Delta \log L_V$
  and $\Delta \log I_e$ for distinguishing different galaxy stellar
  populations at fixed $\sigma$.  {\it Top:} Black (gray) histograms
  show the age and [Fe/H] distributions for ``brighter'' (``fainter'')
  galaxies, as identified by $\Delta \log I_e$.  Galaxies with
  differing stellar populations are well-separated by the $\Delta \log
  I_e$ criterion.  {\it Bottom:} Black (gray) histograms show the age
  and [Fe/H] distributions for brighter (fainter) galaxies, as
  indentified by $\Delta \log L_V$ for the same sample of galaxies
  (data from Paper I).  The $\Delta \log L_V$ criterion also does a
  reasonable job of separating galaxies with different stellar
  populations, but overall provides less clear separation than $\Delta
  \log I_e$: the galaxies in each bin are more mixed (in age and
       [Fe/H]) and the total spread between the peaks (in both age and
       [Fe/H]) is smaller.  Thus $\Delta \log I_e$ is preferable for
       distinguishing galaxies with different star formation histories
       at fixed $\sigma$.  }\label{i_v_lum}
\end{figure}

A similar anti-correlation between residuals in age and Mg abundance
is apparent in Figure \ref{resids}b.  Again, the anti-correlation is
strongly dependent on $\Delta \log I_e$.  The strength of the
anti-correlation is weaker than the age--[Fe/H] trend at fixed
$\sigma$ ($\Delta$[Mg/H] $\propto -0.4 \Delta (\log \mbox{age})$).  This is
consistent with the fact that Figure \ref{sig_v4} shows less spread in
[Mg/H] at fixed $\sigma$ than in [Fe/H].  In addition to the total
abundance of Mg at fixed $\sigma$, there are also systematic
variations in the abundance ratio [Mg/Fe].  Figure \ref{resids}c shows
that [Mg/Fe] is anti-correlated with [Fe/H] at fixed $\sigma$, so that
Fe-poor galaxies are more Mg-enhanced than their Fe-rich counterparts.
As with age, [Fe/H], and [Mg/H], residuals from the $\log
\sigma$--[Mg/Fe] relation are correlated with $\Delta \log I_e$.
Finally, Figure \ref{resids}d shows that residuals in age are
positively correlated with residuals in [Mg/Fe], such that older
galaxies are more Mg-enhanced than younger galaxies at the same
$\sigma$.  The statistical errors in age and [Mg/Fe] are uncorrelated
(i.e., the error ellipses in panel d are horizontal) because, although
age errors in the stellar population modelling process affect
individual abundance measurements, abundance ratio determinations are
robust to modest age errors.  

In summary, there are correlations between the residuals from all of
the various stellar population parameter trends with $\sigma$.  These
depend on $\Delta \log I_e$ such that high-SB galaxies have younger
ages, higher [Fe/H], higher [Mg/H], and lower [Mg/Fe] than low-SB
galaxies with the same $\sigma$ and $R_e$.  These trends are similar
to those reported in Paper I, where more luminous galaxies (those with
higher $\Delta \log L_V$) were similarly younger, more Fe-rich, more
Mg-rich, and with lower [Mg/Fe] with respect to less luminous galaxies
at the same $\sigma$.

In this context, one can ask whether $\Delta \log L_V$ (from Paper I)
or $\Delta \log I_e$ (from this work) is the better discriminator of
differing stellar populations at fixed $\sigma$.  A ``better''
discriminator can be defined as the binning parameter which maximizes
age and [Fe/H] differences at fixed $\sigma$ (i.e., does the best job
of separating galaxies with differing star formation histories from
one another).  Figure \ref{i_v_lum} compares the discriminatory power
of binning galaxies by $\Delta \log I_e$ (as in this analysis) versus
binning by $\Delta \log L_V$ (as in Paper I).  In each analysis,
galaxies were divided into ``bright'', ``medium'', and ``faint'' bins
at fixed $\sigma$, based either on $\Delta \log I_e$ or on $\Delta
\log L_V$.  The ``medium'' brightness bins are not included in Figure
\ref{i_v_lum} because the ``bright'' and ``faint'' bins best
illustrate the total spread in stellar population properties.

The top panels of Figure \ref{i_v_lum} show histograms of the
residuals from the age--$\sigma$ and [Fe/H]--$\sigma$ relations based
on the $\Delta \log I_e$ used in this analysis.  Sorting galaxies by
$\Delta \log I_e$ does an excellent job of separating out galaxies
with differing star formation histories: the galaxies with the high-SB
(black histograms) have the youngest mean ages and the highest [Fe/H],
while those with the low-SB (gray histograms) have the oldest mean
ages and lower [Fe/H].  The separation between the two populations is
very clear.  If $\Delta \log L_V$ is used to bin galaxies at fixed
$\sigma$ (lower panels), the results are generally similar.  However,
the differences between high and low peaks are larger in both cases
with $\Delta \log I_e$ than with $\Delta \log L_V$.  This is as
expected since $\Delta \log I_e$ is a ``pure'' stellar population
measure whereas $\Delta \log L_V$ also contains information from
$\Delta \log R_e$ (at fixed $\sigma$).  As illustrated in Figures
\ref{age_map}--\ref{mgfe_map}, $\Delta \log R_e$ is irrelevant to
stellar population parameters, and thus its presence in $\Delta \log
L_V$ serves only to dilute the information in that quantity.  To
conclude, if one is searching for the best way to identify the
youngest (or most Fe-rich) galaxies at a given $\sigma$, $\Delta \log
I_e$ is a better discriminant than $\Delta \log L_V$.  There may be
cases where radii are not measurable or not available for individual
galaxies, particularly in high redshift surveys, in which case $\Delta
\log L_V$ still provides reasonable discriminatory power.

\begin{figure*}
\begin{center}
\includegraphics[width=0.75\linewidth]{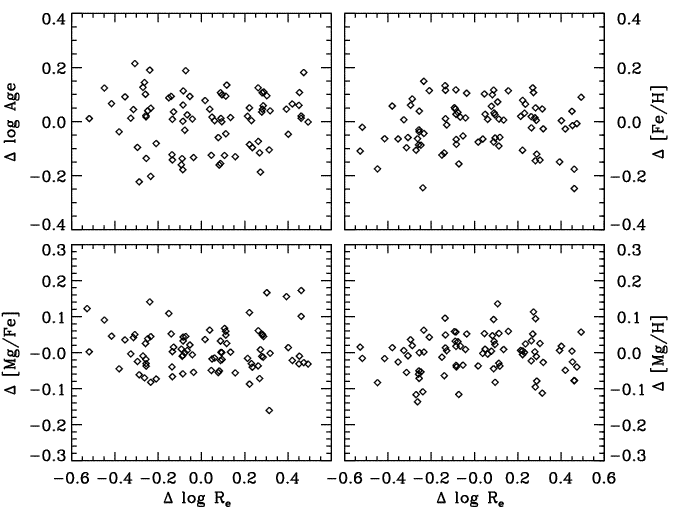}
\caption{Residuals from the mean stellar population trends with
  $\sigma$ as a function of residuals in $R_e$.  The residual $\Delta
  \log R_e$ is computed from the median $\log R_e$ value for each bin
  by subtracting the median $\log R_e$ for {\it all} galaxies at the
  corresponding $\sigma$ (i.e., removing the weak dependence of $R_e$
  on $\sigma$).  As expected from Figures
  \ref{age_map}--\ref{mgfe_map}, there is no correlation between any
  stellar population residual and $\Delta \log R_e$.
}\label{resids_reff}
\end{center}
\end{figure*}

The stellar population maps discussed in \S\ref{mapping} showed very
little variation in stellar population properties as a function of
$R_e$ at fixed $\sigma$.  A direct comparison of stellar population
residuals and their dependence on $R_e$ is shown in Figure
\ref{resids_reff}.  Here, stellar population residuals are shown as a
function of $\Delta \log R_e$, where $\Delta \log R_e$ is defined as
the difference between the median $\log R_e$ for each bin and the
median $\log R_e$ for {\it all} galaxies in the same $\sigma$ range.
The weak dependence of $R_e$ on $\sigma$ has therefore been removed,
highlighting the effect of $R_e$ alone.  It is clear that the stellar
population residuals are unrelated to galaxy size at fixed $\sigma$.

This is an interesting and perhaps surprising result.  The total
$M_{dyn}$ scales as $\sigma^2 R_e$.  Thus, if the fundamental galaxy
property driving stellar population variations were galaxy mass, we
would expect residuals from $\sigma$ trends to correlate with $R_e$ as
well as with $\sigma$.  If this were the case, the contours of age,
[Fe/H], [Mg/H], and [Mg/Fe] in Figures \ref{age_map}--\ref{mgfe_map}
would follow the dashed lines of $M_{dyn}$.  Instead, lines of
constant population properties run nearly vertically in those figures,
and the residuals shown here in Figure \ref{resids_reff} are
completely independent of $R_e$.  This suggests that it is in fact
$\sigma$ and {\it not} total mass that is fundamentally related to the
galaxy star formation history.  The emerging generation of
cosmological models, including predictions for stellar velocity
dispersions, may be able to shed light on the different evolutionary
histories of galaxies with the same total $M_{dyn}$ but different
values of $\sigma$ (and vice versa).

Taken together, these results indicate that $\sigma$ is the primary
structural parameter controlling stellar population variations, and
therefore the primary predictor of a galaxy's past star formation
history.  Indeed, $\sigma$ appears to be more fundamental than total
mass, because residuals from the mean trends between $\sigma$ and the
various stellar population properties do not correlate with $R_e$.  At
fixed $\sigma$, there are additional, genuine correlations between
stellar population parameters which seem to depend strongly on $\Delta
\log I_e$ such that high-SB galaxies are younger, more Fe-rich, and
more Mg-rich, but with less enhanced [Mg/Fe] than typical galaxies at
the same $\sigma$.  Conversely, low-SB galaxies at the same $\sigma$
are older, Fe-poor, Mg-poor, and show significantly enhanced [Mg/Fe].
These correlations provide strong constraints on cosmological models
of galaxy formation and are an important tool for understanding past
star formation histories of early-type galaxies.  The implications for
galaxy star formation histories will be discussed in detail in Paper
III.

\subsection{Can the Observed Variation in Stellar Populations Be
  Caused by Measurement Errors in the FP Parameters?}\label{skeptic}

We have argued that, because the thickness of the FP in $I_e$ is
strongly correlated with stellar population variations, that the
thickness of the plane must be real, rather than merely scatter due to
observational errors.  But before accepting this, it is worth checking
that the observed stellar population trends cannot be {\it caused} by
observational errors in the FP parameters.  It is obvious that, if the
thickness of the FP in the $I_e$ dimension is due to errors in $I_e$
itself, we should not see any trends with $I_e$.  What about thickness
in $I_e$ that is due to errors in $R_e$ and/or $\sigma$ that
``scatter'' galaxies into a different $I_e$ bin?

The slope of the FP indicates that higher $\sigma$ galaxies tend to
have higher $I_e$, thus galaxies with observed high $I_e$ should be
galaxies that have ``scattered'' in from high $\sigma$ (i.e., galaxies
for which $\sigma$ has been underestimated).  However, the galaxies
scattered in from higher $\sigma$ would have older ages and higher
[Mg/Fe], which is opposite to the observed trend with $\Delta \log
I_e$.  It is therefore not possible that errors in $\sigma$ could
produce the observed trends in $I_e$ and stellar population
properties.  In the case of $R_e$, $I_e$ is calculated as $I_e = L / 2
\pi R_e^2$ so overestimating $R_e$ will result in underestimating
$I_e$.  However, we have shown that stellar populations do not depend
on $R_e$ and thus errors in measuring $R_e$ cannot explain the
observed stellar population variations as a function of $I_e$.  This
leads to the conclusion that the thickness of the FP in $I_e$ is
intrinsic to the galaxy population, rather than an artifact of
measurement error, and therefore that the observed stellar population
variation with $\Delta \log I_e$ is likewise real.

\section{Discussion}\label{discussion}

We have seen that, while stellar population properties and therefore
galaxy star formation histories vary substantially with $\sigma$, they
do not vary with $R_e$.  Although this is not a new result (cf.,
\citealt{trager00b}), it is somewhat perplexing, as it implies that
$\sigma$, and {\it not total galaxy mass}, is the better predictor of
galaxy star formation history.  In the literature, $\sigma$ is often
used as a proxy for galaxy mass in stellar population analyses;
authors routinely refer to galaxies with higher $\sigma$ as ``more
massive'', implying that the two parameters are interchangeable.
Various authors have studied stellar population variations as function
of $\sigma$ (e.g., \citealt{trager00b, bernardi03d, thomas05,
smith07-iaus245}), as functions of $M_*$, (e.g.,
\citealt{gallazzi05}), and as functions of $L$ (e.g.,
\citealt{kuntschner98, terlevich02}).  \citet{thomas05} in fact
convert between $\sigma$ and $M_*$ using the mean relation between the
two and present results for stellar population parameters as functions
of either.

Comparisons between all these works are somewhat confusing and do not,
at first glance, appear to give consistent results, particularly when
addressing age variations in galaxies.  This is in part because,
although $\sigma$, $M_*$, and $L$ are all highly correlated, they are
not exactly the same.  It appears that stellar populations vary not
only with $\sigma$, but that at fixed $\sigma$, they vary
systematically with the residuals from the mean $\sigma$--$L$ relation
(as shown in Paper I), so that stellar population trends with $\sigma$
look significantly different from trends with $L$.  (Trends as a
function of $M_*$ are very similar to trends with $L$, as $M_*/L$
varies little among these galaxies.)  We argue in Paper I that the
trends with $\sigma$ are more fundamental and that tracking stellar
population variations with $L$ is not a good proxy for studying them
as a function of $\sigma$.  The work presented here suggests that
$M_{dyn}$ is also not a good proxy for $\sigma$ and confirms that
$\sigma$ is the best predictor of galaxy star formation histories.
Models of galaxy formation must reproduce this dependence on $\sigma$.

Simulations of galaxy mergers offer some insight into why $R_e$ might
be unrelated to galaxy stellar populations.  \citet{robertson06}
present a series of dissipational galaxy merger simulations, along
with a study of how various properties of the mergers affect the FP
parameters of the resulting merger remnant galaxies.  They find that
the final $R_e$ of the merger remnant depends strongly on the
orientation of the initial galaxy orbits and the total angular
momentum of the pre-merger system.  Different initial orientations
produce a factor of two variation in the resulting $R_e$ of the merger
remnant for re-simulations of collisions between identical progenitor
galaxies.  The stellar populations of a galaxy should be independent
of galaxy orbit angular momentum and orientation before a merger;
therefore dissipational mergers will tend to decouple the remnant
$R_e$ from any original relation between galaxy size and star
formation history, should one exist.

Subsequent dissipationless merging of early-type galaxies will also
tend to further scatter galaxies in $R_e$, while having only a small
effect on $\sigma$ and presumably having no effect on the stellar
population of the galaxy.  Simulations of dissipationless mergers
between equal-mass systems by \citet{boylan-kolchin05} result in
remnants with $R_e$ around 70\% (0.23 dex) higher than the pre-merger
systems, while the $\sigma$ of the remnant differs by $< 20$\% (0.06
dex) from the pre-merger system.  Thus equal-mass dissipationless
merging modifies $R_e$ without create major changes in $\sigma$ or the
stellar populations of galaxies, thereby scrambling any pre-existing
correlations.  

These merger simulations suggest that any initial correlation between
galaxy size and star formation history will be erased through
extensive merging.  Gas-rich dissipational mergers will produce
remnants whose radii depend strongly on the orbital orientations of
the progenitors, scrambling any initial size-dependence.  Variable
quantities of subsequent gas-poor dissipationless mergers, which
change $R_e$ but have only minor effects on $\sigma$ and no effect on
the stellar population, will further serve to erase any original
stellar population dependence on $R_e$.

\section{Conclusions}\label{conclusions}

We have identified a sample of $\sim$16,000 early-type galaxies from
the SDSS spectroscopic Main Galaxy Sample to study star formation
histories in FP space.  Bins of similar galaxies are defined along the
FP in $\log \sigma$ and $\log R_e$, and through the thickness of the
FP in $\Delta \log I_e$.  Spectra of the galaxies in each bin are
coadded to produce a set of high $S/N$ mean spectra that span the FP.
The mean spectra are compared with the stellar population models of
\citet{schiavon07} using the automated IDL code {\it EZ\_Ages}
\citep{graves08}, which determines the mean luminosity-weighted age,
abundances [Fe/H], [Mg/H], and the abundance ratio [Mg/Fe] for each
stacked spectrum.  This allows us to study the variations of stellar
populations throughout FP space.

The star formation histories of early-type galaxies are found to form
a two-parameter family.  Interestingly, these two parameters correlate
with $\sigma$ and with $\Delta \log I_e$ but not with $R_e$.  In other
words, the two-parameter family of star formation histories vary
across one dimension of the FP and through the thickness of the FP,
but do not seem to vary with the second dimension across the FP.  A
detailed summary of our results follows.

\begin{list}{}{}
\item[1.]{Stellar population age, [Fe/H], [Mg/H], and [Mg/Fe] all
  increase with increasing galaxy $\sigma$.  This implies a close
  relationship between $\sigma$ and the star formation history of a
  galaxy.  Age and [Fe/H] show substantial spreads at fixed $\sigma$,
  while [Mg/Fe] and [Mg/H] vary less.  The relation between $\sigma$
  and [Mg/H] is particularly strong and tight, showing very little
  spread at high $\sigma$ and only modest spread at low $\sigma$.}
\item[2.]{Stellar population properties are independent of galaxy
  size, $R_e$.  This suggests that $\sigma$ is a better predictor of
  the star formation history of a galaxy than total mass since
  dynamical mass scales as $M_{dyn} \propto \sigma^2 R_e$.  This can
  be understood in the context of merger-driven galaxy formation,
  where $R_e$ is strongly affected by the initial orbital conditions
  of the merging galaxies and may be subsequently further altered by
  dissipationless merging (which does not affect the stellar
  population).}
\item[3.]{At fixed $\sigma$ and $R_e$, stellar populations vary
  substantially with surface brightness residuals from the FP,
  parameterized by $\Delta \log I_e$.  This has a number of
  implications.  Firstly, the thickness of the FP is real rather than
  merely due to measurement error.  Furthermore, the thickness of the
  FP represents an age sequence, such that galaxies with higher
  surface brightness than the midplane of the FP have younger ages,
  while galaxies with lower surface brightness than the midplane have
  older ages.  This result is in agreement with the work of
  \citet{forbes98}, \citet{terlevich02}, and \citet{treu05}.}
\item[4.]{In addition to age variations through the thickness of the
  FP, there are corresponding variations in all other stellar
  population properties studied here.  The variations in age, [Fe/H],
  [Mg/H], and [Mg/Fe] at fixed $\sigma$ are correlated with one
  another such that the galaxies with higher surface brightness than
  the FP midplane are younger, significantly more Fe-rich, only
  slightly more Mg-rich, and have lower [Mg/Fe] than their
  counterparts on the FP.  Similarly, galaxies with lower surface
  brightness than the FP midplane are older, less Fe-rich, slightly
  less Mg-rich, and have substantially enhanced [Mg/Fe] relative to
  their counterparts on the FP.  The general age-metallicity
  anti-correlation is consistent with previous work by
  \citet{trager00b} and \citet{smith07-iaus245}.  Although these
  co-variations are in the directions of the correlated errors in
  stellar population modelling, neither the statistical errors nor the
  known systematic errors in the modelling process can explain the
  variations, strongly suggesting that they are real effects.  }
\item[5.]{The variations with $\Delta \log I_e$ at fixed $\sigma$ and
  $R_e$ demonstrated in this analysis are consistent with the results
  of Paper I, which showed similar stellar population variations with
  $\Delta \log L_V$ at fixed $\sigma$.  However, the FP residual
  $\Delta \log I_e$ does a somewhat better job of distinguishing young
  Fe-rich populations from old Fe-poor populations at the same
  $\sigma$ and is thus preferable to $\Delta \log L_V$ for studying
  variations in the star formation histories of early-type galaxies.
}
\end{list}

This work demonstrates that the stellar populations of early-type
galaxy form a two-parameter family.  This two-parameter family maps
onto a cross-section through the FP.  Paper III in this series will
explicitly quantify this mapping, as well as translating the derived
stellar population parameters into model star formation histories.

\acknowledgements

The authors would like to thank Arjen van der Wel for providing
effective radius measurements based on GALFIT photometry as a check on
the SDSS pipeline radii, and Renbin Yan for providing the emission
line measurements used to identify the sample of early-type galaxies
used here.  This work was supported by National Science Foundation
grant AST 05-07483.  

Funding for the creation and distribution of the SDSS Archive has been
provided by the Alred P. Sloan Foundation, the Participating
Institutions, the National Aeronautics and Space Administration, the
National Science Foundation, the US Department of Energy, the Japanese
Monbukagakusho, and the Max-Planck Society. The SDSS Web site is
http://www.sdss.org/.

The SDSS is managed by the Astrophysical Research Consortium (ARC) for
the Participating Institutions. The Participating Institutions are the
University of Chicago, Fermilab, the Institute for Advanced Study, the
Japan Participation Group, the Johns Hopkins University, the Korean
Scientist Group, Los Alamos National Laboratory, the
Max-Planck-Institute for Astronomy (MPIA), the Max-Planck-Institute
for Astrophysics (MPA), New Mexico State University, University of
Pittsburgh, University of Portsmouth, Princeton University, the United
States Naval Observatory, and the University of Washington.

\appendix

\section{Frosting Models}\label{frosting}

Section \ref{residuals} showed that correlated statistical errors
cannot account for the co-variation of stellar population properties
with $\Delta \log I_e$ at fixed $\sigma$.  The younger ages observed
for high-SB galaxies are genuine variations, in agreement with the
results of \citet{forbes98}.  However, the stellar population models
used in this work are single-burst models.  Because young stars are
luminous and contribute disproportionately to integrated galaxy light
and Balmer absorption line strengths, a small population of young
stars can skew the ``mean'' age to younger values than a true,
mass-weighted mean age.  The effect is largest when the age difference
between sub-populations is large compared to the mean stellar age, so
that the youngest age measurements in our data are likely to be
somewhat too young.  Determinations of age and [Fe/H] from absorption
indices are anti-correlated, so a young component would not only skew
the mean age to younger values but should also skew the measured
[Fe/H] to higher values.  Because [Mg/Fe] is an abundance ratio, it is
relatively insensitive to the correlated errors in age and [Fe/H] (see
Figure 3 of \citet{graves08}.  The abundance [Mg/H] is calculated
as [Fe/H] + [Mg/Fe], so errors in [Fe/H] will translate directly into
comparable errors in [Mg/H].

\begin{figure}[t]
\begin{center}
\includegraphics[width=0.7\linewidth]{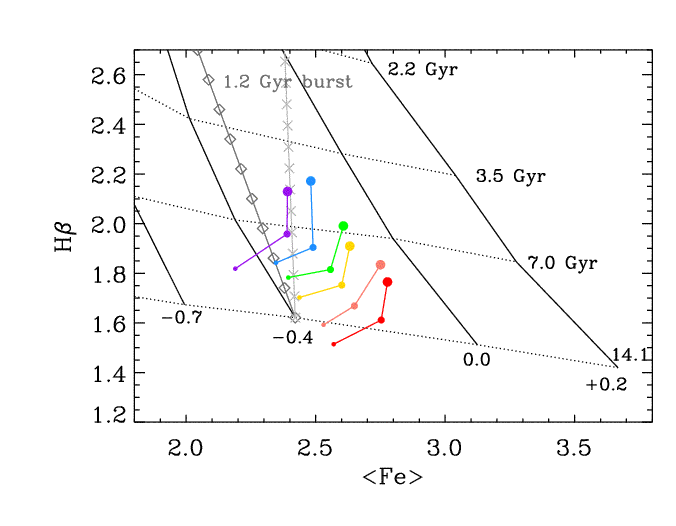}
\caption{Comparing observed index variations with two-burst
  ``frosting'' stellar population models.  The black grid shows
  solar-ratio single burst models from \citet{schiavon07}.  Solid
  lines indicate lines of constant [Fe/H], while dotted lines indicate
  lines of constant age, as labelled.  The dark gray trajectory shows
  the effect of adding a 1.2 Gyr burst with [Fe/H] = -0.4 to an
  underlying 14.1 Gyr old population with the same [Fe/H].  Diamonds
  indicate the fraction of the galaxy light at $\sim5000$ {\AA}
  contributed by the young population, starting from 0\% at the bottom
  and increasing in increments of 5\% to 45\% at the top of the
  diagram.  The light gray trajectory shows the effect of adding a 1.2
  Gyr burst with four times larger Fe abundance ([Fe/H] = +0.2) to the
  underlying 14.1 Gyr population.  ``X'''s indicate the fraction of
  light contributed by the young Fe-rich population in increments of
  5\%.  The indices measured in our stacked SDSS spectra are shown as
  colored points, with purple, blue, green, gold, orange, and red
  indicating the lowest through highest $\sigma$ bins, respectively.
  Small, medium, and large circles indicate the low-SB, midplane, and
  high-SB bins in $\Delta \log I_e$, respectively.  Differences
  between the low-SB and midplane bins cannot be produced merely by
  adding a young subpopulation of stars---a genuinely higher [Fe/H]
  population is required.  Differences between the midplane and
  high-SB bins are compatible with a young subpopulation at the same
  [Fe/H] for the lowest $\sigma$ bin only---higher $\sigma$ bins also
  require a genuinely higher [Fe/H] population.  The observed
  anti-correlation between mean age and [Fe/H] at fixed $\sigma$
  therefore reflects genuine differences in the stellar population
  abundances and cannot be caused solely by systematic effects from
  using single burst stellar population models.  }\label{burst_grid}
\end{center}
\end{figure}

Figure \ref{burst_grid} shows absorption line strength predictions for
H$\beta$ and $\langle$Fe$\rangle$ from stellar population models.  The
black grid shows a set of single-burst models from \citet{schiavon07}
with solar abundance ratios.  Solid lines indicate lines of constant
[Fe/H], while dotted lines indicate lines of constant age, as labelled
in the figure.  In addition to the grid of single burst models, two
different frosting models are shown.  The dark gray line shows the
effect of adding a burst with age 1.2 Gyr to an underlying stellar
population with an age of 14.1 Gyr.  Both the old and young population
components have [Fe/H]$ = -0.4$.  Diamonds indicate the fraction of
galaxy light contributed by the young burst, starting at the bottom at
0\% and increasing in increments of 5\% moving upward along the line.
This line shows the trajectory in H$\beta$--$\langle$Fe$\rangle$ space
produced by adding young bursts at the same [Fe/H] to an underlying
old population.  Though this model is computed for a burst, the
direction traced by adding a {\it range} of younger stars would be the
same or tilted even more strongly to the left.  The light gray line
with ``X'' symbols shows a similar frosting model, but here the young
1.2 Gyr burst is four times more Fe-rich than the underlying old
population, with [Fe/H]$= +0.2$.

The colored points show the data from our stacked galaxy spectra.
Different colors represent the different bins in $\sigma$, with purple
indicating the lowest-$\sigma$ bin and red indicating the
highest-$\sigma$ bin.  At each value of $\sigma$, the three circles
indicate the three bins in $\Delta \log I_e$, with the smallest circle
showing the low-SB bin and the largest circle showing the high-SB
bin.  The bins have been averaged over $R_e$.

It is interesting that the slopes between the low-SB and midplane bins
(small and medium circles) are substantially different from those
between the midplane and high-SB bins (medium and large circles).
Comparing first the low-SB and midplane bins, it is clear that the
differences between low-SB and midplane galaxies cannot be explained at
all by a small frosting of younger stars at the same [Fe/H].  The
same-Fe frosting model (dark gray line with diamonds) is nearly
orthogonal to the change in H$\beta$--$\langle$Fe$\rangle$ space
between the low-SB and midplane bins.  Even adding a frosting
population with four times larger Fe abundance (light gray lines with
exes) does not come close to explaining the $\langle$Fe$\rangle$ line
strength differences between the low-SB and midplane galaxies.  The
low-SB galaxies must have genuinely lower values of [Fe/H] than the
galaxies at the FP midplane, as well as having older ages.

Line strengths differences between the midplane galaxies and the
high-SB galaxies are more similar to the
H$\beta$--$\langle$Fe$\rangle$ trajectories expected for frosting
models, particularly for low-$\sigma$ galaxies.  At high $\sigma$,
neither the constant [Fe/H] nor the Fe-rich frosting models can
account for the observed increases in $\langle$Fe$\rangle$ in the
high-SB galaxies.  For the lowest $\sigma$ bin, the difference between
the midplane galaxies and the high-SB galaxies is consistent with a
$\sim 5$\% frosting of young stars at similar [Fe/H] to the underlying
population---no genuine difference in [Fe/H] is needed to explain the
data.  However, the difference in [Fe/H] inferred from assuming a
single burst population is only 0.04 dex.  This indicates that the
simplifying assumption of a single burst population only biases the
estimated [Fe/H] by a small amount for the high-SB, lowest-$\sigma$
galaxies, where the bias is strongest.  This small bias in the
measured [Fe/H] will translate into a similarly small bias in [Mg/H].

In general, a small frosting of younger stars, while adequate to
explain the observed differences in stellar population ages at fixed
$\sigma$, cannot also produce the observed variation in [Fe/H] and
[Mg/H].  The variation in $\langle$Fe$\rangle$ linestrength at fixed
$\sigma$ requires a real variation in [Fe/H].  Thus the variations in
observed abundance patterns at fixed $\sigma$ must be due to genuine
abundance differences between the galaxies, rather than an artifact of
the modelling process for the younger galaxies.

\bibliographystyle{apj}
\bibliography{apj-jour,myrefs}

\begin{thebibliography}{44}
\expandafter\ifx\csname natexlab\endcsname\relax\def\natexlab#1{#1}\fi

\bibitem[{{Adelman-McCarthy} {et~al.}(2006)}]{adelman-mccarthy06}
{Adelman-McCarthy}, J.~K., et~al. 2006, \apjs, 162, 38

\bibitem[{{Adelman-McCarthy}(2008)}]{adelman-mccarthy08}
{Adelman-McCarthy}, J.~K.~f. 2008, \apjs, 175, 297 

\bibitem[{{Bender} {et~al.}(1992){Bender}, {Burstein}, \& {Faber}}]{bender92}
{Bender}, R., {Burstein}, D., \& {Faber}, S.~M. 1992, \apj, 399, 462

\bibitem[{{Bernardi} {et~al.}(2003{\natexlab{a}})}]{bernardi03c}
{Bernardi}, M., et~al. 2003{\natexlab{a}}, \aj, 125, 1866

\bibitem[{{Bernardi} {et~al.}(2003{\natexlab{b}})}]{bernardi03d}
  ---. 2003{\natexlab{b}}, \aj, 125, 1882

\bibitem[{{Blanton} {et~al.}(2003)}]{blanton03-kcorrect} {Blanton},
M.~R., et~al. 2003, \aj, 125, 2348

\bibitem[{{Blanton} {et~al.}(2005)}]{blanton05-vagc} {Blanton}, M.~R.,
et~al. 2005, \aj, 129, 2562

\bibitem[{{Boylan-Kolchin} {et~al.}(2005){Boylan-Kolchin}, {Ma}, \&
  {Quataert}}]{boylan-kolchin05}
{Boylan-Kolchin}, M., {Ma}, C.-P., \& {Quataert}, E. 2005, \mnras, 362, 184

\bibitem[{{Cardiel} {et~al.}(1998){Cardiel}, {Gorgas}, {Cenarro}, \&
  {Gonzalez}}]{cardiel98}
{Cardiel}, N., {Gorgas}, J., {Cenarro}, J., \& {Gonzalez}, J.~J. 1998, \aaps,
  127, 597

\bibitem[{{Djorgovski} \& {Davis}(1987)}]{djorgovski87}
{Djorgovski}, S. \& {Davis}, M. 1987, \apj, 313, 59

\bibitem[{{Dressler} {et~al.}(1987){Dressler}, {Lynden-Bell}, {Burstein},
  {Davies}, {Faber}, {Terlevich}, \& {Wegner}}]{dressler87}
{Dressler}, A., {Lynden-Bell}, D., {Burstein}, D., {Davies}, R.~L., {Faber},
  S.~M., {Terlevich}, R., \& {Wegner}, G. 1987, \apj, 313, 42

\bibitem[{{Faber} {et~al.}(1987){Faber}, {Dressler}, {Davies}, {Burstein}, \&
  {Lynden-Bell}}]{faber87}
{Faber}, S.~M., {Dressler}, A., {Davies}, R.~L., {Burstein}, D., \&
  {Lynden-Bell}, D. 1987, in Nearly Normal Galaxies. From the Planck Time to
  the Present, ed. S.~M. {Faber}, 175--183

\bibitem[{{Fisher} \& {Drory}(2008)}]{fisher08}
{Fisher}, D.~B. \& {Drory}, N. 2008, \aj, 136, 773

\bibitem[{{Forbes} {et~al.}(1998){Forbes}, {Ponman}, \& {Brown}}]{forbes98}
{Forbes}, D.~A., {Ponman}, T.~J., \& {Brown}, R.~J.~N. 1998, \apjl, 508, L43

\bibitem[{{Gallazzi} {et~al.}(2005){Gallazzi}, {Charlot}, {Brinchmann},
  {White}, \& {Tremonti}}]{gallazzi05}
{Gallazzi}, A., {Charlot}, S., {Brinchmann}, J., {White}, S.~D.~M., \&
  {Tremonti}, C.~A. 2005, \mnras, 362, 41

\bibitem[{{Graves} {et~al.}(2009){Graves}, {Faber}, \&
  Schiavon}]{graves08_paperI}
{Graves}, G.~J., {Faber}, S.~M., \& Schiavon, R.~P. 2009, \apj, 693, 486

\bibitem[{{Graves} {et~al.}(2007){Graves}, {Faber}, {Schiavon}, \&
  {Yan}}]{graves07}
{Graves}, G.~J., {Faber}, S.~M., {Schiavon}, R.~P., \& {Yan}, R. 2007, \apj,
  671, 243

\bibitem[{{Graves} \& Schiavon(2008)}]{graves08}
{Graves}, G.~J. \& Schiavon, R.~P. 2008, \apjs, 177, 446

\bibitem[{{J{\o}rgensen} {et~al.}(2006){J{\o}rgensen}, {Chiboucas}, {Flint},
  {Bergmann}, {Barr}, \& {Davies}}]{jorgensen06}
{J{\o}rgensen}, I., {Chiboucas}, K., {Flint}, K., {Bergmann}, M., {Barr}, J.,
  \& {Davies}, R. 2006, \apjl, 639, L9

\bibitem[{{J{\o}rgensen} {et~al.}(1996){J{\o}rgensen}, {Franx}, \&
  {Kjaergaard}}]{jorgensen96}
{J{\o}rgensen}, I., {Franx}, M., \& {Kjaergaard}, P. 1996, \mnras, 280, 167

\bibitem[{{Kauffmann} {et~al.}(2003)}]{kauffmann03} {Kauffmann}, G.,
et~al. 2003, \mnras, 346, 1055

\bibitem[{{Kormendy} \& {Fisher}(2008)}]{kormendy08}
{Kormendy}, J. \& {Fisher}, D.~B. 2008, in Astronomical Society of the Pacific
  Conference Series, Vol. 396, Astronomical Society of the Pacific Conference
  Series, ed. J.~G. {Funes} \& E.~M. {Corsini}, 297

\bibitem[{{Kuntschner} \& {Davies}(1998)}]{kuntschner98}
{Kuntschner}, H. \& {Davies}, R.~L. 1998, \mnras, 295, L29

\bibitem[{{Nelan} {et~al.}(2005){Nelan}, {Smith}, {Hudson}, {Wegner}, {Lucey},
  {Moore}, {Quinney}, \& {Suntzeff}}]{nelan05}
{Nelan}, J.~E., {Smith}, R.~J., {Hudson}, M.~J., {Wegner}, G.~A., {Lucey},
  J.~R., {Moore}, S.~A.~W., {Quinney}, S.~J., \& {Suntzeff}, N.~B. 2005, \apj,
  632, 137

\bibitem[{{Peng} {et~al.}(2002){Peng}, {Ho}, {Impey}, \& {Rix}}]{peng02}
{Peng}, C.~Y., {Ho}, L.~C., {Impey}, C.~D., \& {Rix}, H.-W. 2002, \aj, 124, 266

\bibitem[{{Robertson} {et~al.}(2006){Robertson}, {Cox}, {Hernquist}, {Franx},
  {Hopkins}, {Martini}, \& {Springel}}]{robertson06}
{Robertson}, B., {Cox}, T.~J., {Hernquist}, L., {Franx}, M., {Hopkins}, P.~F.,
  {Martini}, P., \& {Springel}, V. 2006, \apj, 641, 21

\bibitem[{{Schawinski} {et~al.}(2007){Schawinski}, {Thomas}, {Sarzi},
  {Maraston}, {Kaviraj}, {Joo}, {Yi}, \& {Silk}}]{schawinski07}
{Schawinski}, K., {Thomas}, D., {Sarzi}, M., {Maraston}, C., {Kaviraj}, S.,
  {Joo}, S.-J., {Yi}, S.~K., \& {Silk}, J. 2007, \mnras, 382, 1415

\bibitem[{{Schiavon}(2007)}]{schiavon07}
{Schiavon}, R.~P. 2007, \apjs, 171, 146

\bibitem[{{Sersic}(1968)}]{sersic68}
{Sersic}, J.~L. 1968, {Atlas de galaxias australes} (Cordoba, Argentina:
  Observatorio Astronomico, 1968)

\bibitem[{{Smith} {et~al.}(2007{\natexlab{a}}){Smith}, {Lucey}, \&
  {Hudson}}]{smith07}
{Smith}, R.~J., {Lucey}, J.~R., \& {Hudson}, M.~J. 2007{\natexlab{a}}, \mnras,
  381, 1035

\bibitem[{{Smith} {et~al.}(2008{\natexlab{b}}){Smith}, {Lucey}, \&
  {Hudson}}]{smith07-iaus245} ---. 2008{\natexlab{b}}, in Proc. of
  IAUS 245, Formation and Evolution of Galaxy Bulges, p. 411-414,
  arXiv:0712.0274

\bibitem[{{Strauss} {et~al.}(2002)}]{strauss02} {Strauss}, M.~A.,
et~al. 2002, \aj, 124, 1810

\bibitem[{{Terlevich} \& {Forbes}(2002)}]{terlevich02}
{Terlevich}, A.~I. \& {Forbes}, D.~A. 2002, \mnras, 330, 547

\bibitem[{{Thomas} {et~al.}(2005){Thomas}, {Maraston}, {Bender}, \& {Mendes de
  Oliveira}}]{thomas05}
{Thomas}, D., {Maraston}, C., {Bender}, R., \& {Mendes de Oliveira}, C. 2005,
  \apj, 621, 673

\bibitem[{{Trager} {et~al.}(2000){Trager}, {Faber}, {Worthey}, \&
  {Gonz{\'a}lez}}]{trager00b}
{Trager}, S.~C., {Faber}, S.~M., {Worthey}, G., \& {Gonz{\'a}lez}, J.~J. 2000,
  \aj, 120, 165

\bibitem[{{Treu} {et~al.}(2005{\natexlab{a}}){Treu}, {Ellis}, {Liao}, \& {van
  Dokkum}}]{treu05-apjl}
{Treu}, T., {Ellis}, R.~S., {Liao}, T.~X., \& {van Dokkum}, P.~G.
  2005{\natexlab{a}}, \apjl, 622, L5

\bibitem[{{Treu} {et~al.}(2005{\natexlab{b}})}]{treu05} {Treu}, T.,
et~al. 2005{\natexlab{b}}, \apj, 633, 174

\bibitem[{{van der Wel} {et~al.}(2004){van der Wel}, {Franx}, {van Dokkum}, \&
  {Rix}}]{vanderwel04}
{van der Wel}, A., {Franx}, M., {van Dokkum}, P.~G., \& {Rix}, H.-W. 2004,
  \apjl, 601, L5

\bibitem[{{van der Wel} {et~al.}(2005){van der Wel}, {Franx}, {van Dokkum},
  {Rix}, {Illingworth}, \& {Rosati}}]{vanderwel05}
{van der Wel}, A., {Franx}, M., {van Dokkum}, P.~G., {Rix}, H.-W.,
  {Illingworth}, G.~D., \& {Rosati}, P. 2005, \apj, 631, 145

\bibitem[{{Worthey} {et~al.}(1994){Worthey}, {Faber}, {Gonzalez}, \&
  {Burstein}}]{worthey94}
{Worthey}, G., {Faber}, S.~M., {Gonzalez}, J.~J., \& {Burstein}, D. 1994,
  \apjs, 94, 687

\bibitem[{{Worthey} \& {Ottaviani}(1997)}]{worthey97}
{Worthey}, G. \& {Ottaviani}, D.~L. 1997, \apjs, 111, 377

\bibitem[{{Worthey} {et~al.}(1995){Worthey}, {Trager}, \& {Faber}}]{worthey95}
{Worthey}, G., {Trager}, S.~C., \& {Faber}, S.~M. 1995, in Astronomical Society
  of the Pacific Conference Series, Vol.~86, Fresh Views of Elliptical
  Galaxies, ed. A.~{Buzzoni}, A.~{Renzini}, \& A.~{Serrano}, 203

\bibitem[{{Yan} {et~al.}(2006){Yan}, {Newman}, {Faber}, {Konidaris}, {Koo}, \&
  {Davis}}]{yan06}
{Yan}, R., {Newman}, J.~A., {Faber}, S.~M., {Konidaris}, N., {Koo}, D., \&
  {Davis}, M. 2006, \apj, 648, 281

\bibitem[{{York} {et~al.}(2000)}]{york00} {York}, D.~G., et~al. 2000,
\aj, 120, 1579

\end{thebibliography}

\clearpage

\setcounter{table}{0} \renewcommand{\thetable}{\arabic{table}} 

\LongTables
\begin{deluxetable*}{cccr@{ $< \Delta I_e <$ }l@{}cccccc}
\tabletypesize{\scriptsize}
\tablecaption{Properties of Galaxy Bins and Stacked Spectra\label{bin_tab}}
\tablewidth{0pt}
\tablehead{
\colhead{} &
\colhead{} &
\colhead{} &
\multicolumn{2}{c}{} &
\colhead{Median} &
\colhead{Median} &
\colhead{Median} &
\colhead{Median} &
\colhead{} &
\colhead{} \\
\colhead{} &
\colhead{$\sigma$ bin\tablenotemark{*}} &
\colhead{$R_e$ bin\tablenotemark{*}} &
\multicolumn{2}{c}{$I_e$ bin\tablenotemark{*}} &
\colhead{$\log \sigma$} &
\colhead{$\log R_e$} &
\colhead{$\Delta \log I_e$} &
\colhead{$\log I_e$} &
\colhead{Number\tablenotemark{\dag}} &
\colhead{$S/N$\tablenotemark{\ddag}} \\
\colhead{} &
\colhead{(km s$^{-1}$)} &
\colhead{(kpc)} &
\multicolumn{2}{c}{($L_{\odot}$ pc$^{-2}$)} &
\colhead{(km s$^{-1}$)} &
\colhead{(kpc)} &
\colhead{($L_{\odot}$ pc$^{-2}$)} &
\colhead{($L_{\odot}$ pc$^{-2}$)} &
\colhead{} &
\colhead{({\AA}$^{-1}$)}
}
\startdata
 1 &$1.86 < \sigma < 2.00$ &-0.1 $< R_e < $ 0.1 &-0.3 &-0.1 &1.96 &0.05 &-0.17 &2.60 & 33 & 85 \\    
 2 &	    		   &                    &-0.1 &\phantom{-}0.1 &1.94 &0.04 &\phantom{-}0.01 &2.78 & 86 &158 \\
 3 &	    		   &                    & 0.1 &\phantom{-}0.3 &1.94 &0.05 &\phantom{-}0.19 &2.95 & 39 &115 \\ 
 4 &	    		   &0.1  $< R_e < $ 0.3 &-0.3 &-0.1 &1.95 &0.21 &-0.17 &2.37 & 89 &154 \\
 5 &	    		   &                    &-0.1 &\phantom{-}0.1 &1.95 &0.22 &\phantom{-}0.00 &2.55 &276 &273 \\
 6 &	    		   &                    & 0.1 &\phantom{-}0.3 &1.94 &0.20 &\phantom{-}0.15 &2.72 &133 &214 \\
 7 &	    		   &0.3  $< R_e < $ 0.5 &-0.3 &-0.1 &1.95 &0.40 &-0.16 &2.15 &119 &159 \\
 8 &	    		   &                    &-0.1 &\phantom{-}0.1 &1.94 &0.40 &-0.01 &2.32 &263 &259 \\
 9 &	    		   &                    & 0.1 &\phantom{-}0.3 &1.95 &0.39 &\phantom{-}0.14 &2.48 & 65 &154 \\
10 &	    		   &0.5  $< R_e < $ 0.7 &-0.3 &-0.1 &1.96 &0.55 &-0.17 &1.97 & 36 & 80 \\
11 &	    		   &                    &-0.1 &\phantom{-}0.1 &1.94 &0.57 &-0.01 &2.10 & 75 &152 \\
12 &	    		   &                    & 0.1 &\phantom{-}0.3 &1.93 &0.55 &\phantom{-}0.13 &2.26 & 23 & 89 \\
13 &	    		   &0.7  $< R_e < $ 0.9 &-0.3 &-0.1 &...  &...  &...   &...  &  3 & 27 \\
14 &	    		   &                    &-0.1 &\phantom{-}0.1 &1.96 &0.76 &\phantom{-}0.01 &1.86 & 18 & 77 \\
15 &                       &                    & 0.1 &\phantom{-}0.3 &...  &...  &...   &...  &  5 & 41 \\
16 &$2.00 < \sigma < 2.09$ &-0.1 $< R_e < $ 0.1 &-0.3 &-0.1 &2.05 &0.03 &-0.15 &2.75 & 51 &130 \\
17 &                       &                    &-0.1 &\phantom{-}0.1 &2.06 &0.03 &\phantom{-}0.01 &2.91 &190 &262 \\
18 &                       &                    & 0.1 &\phantom{-}0.3 &2.06 &0.00 &\phantom{-}0.15 &3.09 & 73 &180 \\
19 &                       &0.1  $< R_e < $ 0.3 &-0.3 &-0.1 &2.05 &0.24 &-0.15 &2.50 &157 &207 \\
20 &                       &                    &-0.1 &\phantom{-}0.1 &2.06 &0.21 &-0.00 &2.68 &491 &419 \\
21 &                       &                    & 0.1 &\phantom{-}0.3 &2.05 &0.20 &\phantom{-}0.15 &2.86 &108 &225 \\
22 &                       &0.3  $< R_e < $ 0.5 &-0.3 &-0.1 &2.05 &0.38 &-0.15 &2.30 &186 &224 \\
23 &                       &                    &-0.1 &\phantom{-}0.1 &2.05 &0.38 &-0.01 &2.46 &470 &421 \\
24 &                       &                    & 0.1 &\phantom{-}0.3 &2.06 &0.38 &\phantom{-}0.15 &2.65 & 93 &221 \\
25 &                       &0.5  $< R_e < $ 0.7 &-0.3 &-0.1 &2.04 &0.58 &-0.17 &2.04 & 64 &120 \\
26 &                       &                    &-0.1 &\phantom{-}0.1 &2.06 &0.57 &-0.01 &2.23 &131 &228 \\
27 &                       &                    & 0.1 &\phantom{-}0.3 &2.06 &0.56 &\phantom{-}0.15 &2.44 & 20 &119 \\
28 &                       &0.7  $< R_e < $ 0.9 &-0.3 &-0.1 &2.05 &0.75 &-0.12 &1.83 & 12 & 56 \\
29 &                       &                    &-0.1 &\phantom{-}0.1 &2.06 &0.74 &\phantom{-}0.02 &2.02 & 24 &101 \\
30 &                       &                    & 0.1 &\phantom{-}0.3 &...  &...  &...   &...  &  2 & 31 \\
31 &$2.09 < \sigma < 2.18$ &-0.1 $< R_e < $ 0.1 &-0.3 &-0.1 &2.14 &0.03 &-0.16 &2.84 &100 &190 \\
32 &                       &                    &-0.1 &\phantom{-}0.1 &2.14 &0.04 &\phantom{-}0.00 &3.00 &348 &371 \\
33 &                       &                    & 0.1 &\phantom{-}0.3 &2.13 &0.04 &\phantom{-}0.14 &3.15 &124 &257 \\
34 &                       &0.1  $< R_e < $ 0.3 &-0.3 &-0.1 &2.13 &0.21 &-0.15 &2.62 &240 &274 \\
35 &                       &                    &-0.1 &\phantom{-}0.1 &2.14 &0.21 &-0.00 &2.78 &817 &594 \\
36 &                       &                    & 0.1 &\phantom{-}0.3 &2.14 &0.21 &\phantom{-}0.15 &2.95 &190 &355 \\
37 &                       &0.3  $< R_e < $ 0.5 &-0.3 &-0.1 &2.13 &0.38 &-0.14 &2.41 &254 &286 \\
38 &                       &                    &-0.1 &\phantom{-}0.1 &2.14 &0.39 &-0.00 &2.56 &785 &644 \\
39 &                       &                    & 0.1 &\phantom{-}0.3 &2.14 &0.38 &\phantom{-}0.15 &2.73 &189 &376 \\
40 &                       &0.5  $< R_e < $ 0.7 &-0.3 &-0.1 &2.14 &0.58 &-0.16 &2.15 & 81 &165 \\
41 &                       &                    &-0.1 &\phantom{-}0.1 &2.14 &0.58 &-0.00 &2.33 &343 &439 \\
42 &                       &                    & 0.1 &\phantom{-}0.3 &2.15 &0.57 &\phantom{-}0.15 &2.51 & 77 &250 \\
43 &                       &0.7  $< R_e < $ 0.9 &-0.3 &-0.1 &2.14 &0.76 &-0.16 &1.92 & 13 & 61 \\
44 &                       &                    &-0.1 &\phantom{-}0.1 &2.14 &0.75 &\phantom{-}0.02 &2.13 & 60 &174 \\
45 &                       &                    & 0.1 &\phantom{-}0.3 &2.15 &0.79 &\phantom{-}0.13 &2.24 & 24 &132 \\
46 &$2.18 < \sigma < 2.27$ &-0.1 $< R_e < $ 0.1 &-0.3 &-0.1 &2.23 &0.04 &-0.15 &2.94 & 96 &197 \\
47 &                       &                    &-0.1 &\phantom{-}0.1 &2.22 &0.03 &-0.01 &3.10 &302 &387 \\
48 &                       &                    & 0.1 &\phantom{-}0.3 &2.21 &0.05 &\phantom{-}0.14 &3.23 & 66 &200 \\
49 &                       &0.1  $< R_e < $ 0.3 &-0.3 &-0.1 &2.22 &0.21 &-0.15 &2.73 &213 &286 \\
50 &                       &                    &-0.1 &\phantom{-}0.1 &2.22 &0.21 &-0.01 &2.88 &679 &630 \\
51 &                       &                    & 0.1 &\phantom{-}0.3 &2.22 &0.21 &\phantom{-}0.16 &3.04 &186 &388 \\
52 &                       &0.3  $< R_e < $ 0.5 &-0.3 &-0.1 &2.22 &0.39 &-0.14 &2.51 &175 &293 \\
53 &                       &                    &-0.1 &\phantom{-}0.1 &2.22 &0.40 &\phantom{-}0.01 &2.66 &871 &779 \\
54 &                       &                    & 0.1 &\phantom{-}0.3 &2.23 &0.39 &\phantom{-}0.15 &2.82 &227 &463 \\
55 &                       &0.5  $< R_e < $ 0.7 &-0.3 &-0.1 &2.23 &0.57 &-0.15 &2.29 & 81 &192 \\
56 &                       &                    &-0.1 &\phantom{-}0.1 &2.23 &0.58 &\phantom{-}0.01 &2.44 &555 &637 \\
57 &                       &                    & 0.1 &\phantom{-}0.3 &2.23 &0.58 &\phantom{-}0.15 &2.57 &118 &337 \\
58 &                       &0.7  $< R_e < $ 0.9 &-0.3 &-0.1 &2.21 &0.74 &-0.14 &2.08 & 23 & 97 \\
59 &                       &                    &-0.1 &\phantom{-}0.1 &2.23 &0.77 &\phantom{-}0.01 &2.22 &125 &294 \\
60 &                       &                    & 0.1 &\phantom{-}0.3 &2.25 &0.75 &\phantom{-}0.14 &2.38 & 36 &183 \\
61 &$2.27 < \sigma < 2.36$ &-0.1 $< R_e < $ 0.1 &-0.3 &-0.1 &2.31 &0.00 &-0.14 &3.06 & 49 &145 \\
62 &                       &                    &-0.1 &\phantom{-}0.1 &2.30 &0.04 &-0.03 &3.19 &127 &273 \\
63 &                       &                    & 0.1 &\phantom{-}0.3 &2.30 &0.07 &\phantom{-}0.14 &3.30 & 12 &101 \\
64 &                       &0.1  $< R_e < $ 0.3 &-0.3 &-0.1 &2.30 &0.19 &-0.15 &2.83 & 99 &229 \\
65 &                       &                    &-0.1 &\phantom{-}0.1 &2.30 &0.22 &-0.02 &2.95 &278 &451 \\
66 &                       &                    & 0.1 &\phantom{-}0.3 &2.30 &0.24 &\phantom{-}0.13 &3.08 & 53 &235 \\
67 &                       &0.3  $< R_e < $ 0.5 &-0.3 &-0.1 &2.31 &0.37 &-0.14 &2.63 &111 &268 \\
68 &                       &                    &-0.1 &\phantom{-}0.1 &2.30 &0.41 &-0.01 &2.73 &527 &688 \\
69 &                       &                    & 0.1 &\phantom{-}0.3 &2.30 &0.42 &\phantom{-}0.15 &2.89 &106 &349 \\
70 &                       &0.5  $< R_e < $ 0.7 &-0.3 &-0.1 &2.32 &0.56 &-0.13 &2.42 & 54 &186 \\
71 &                       &                    &-0.1 &\phantom{-}0.1 &2.31 &0.59 &\phantom{-}0.01 &2.53 &504 &675 \\
72 &                       &                    & 0.1 &\phantom{-}0.3 &2.31 &0.61 &\phantom{-}0.14 &2.65 &161 &422 \\
73 &                       &0.7  $< R_e < $ 0.9 &-0.3 &-0.1 &2.33 &0.75 &-0.12 &2.19 & 11 & 89 \\
74 &                       &                    &-0.1 &\phantom{-}0.1 &2.32 &0.77 &\phantom{-}0.03 &2.33 &213 &442 \\
75 &                       &                    & 0.1 &\phantom{-}0.3 &2.32 &0.77 &\phantom{-}0.14 &2.45 & 67 &278 \\
76 &$2.36 < \sigma < 2.50$ &-0.1 $< R_e < $ 0.1 &-0.3 &-0.1 &2.38 &0.02 &-0.15 &3.17 & 19 &100 \\
77 &                       &                    &-0.1 &\phantom{-}0.1 &2.40 &0.03 &-0.03 &3.28 & 37 &168 \\
78 &                       &                    & 0.1 &\phantom{-}0.3 &...  &...  &...   &...  &  3 & 53 \\
79 &                       &0.1  $< R_e < $ 0.3 &-0.3 &-0.1 &2.39 &0.19 &-0.15 &2.94 & 32 &149 \\
80 &                       &                    &-0.1 &\phantom{-}0.1 &2.38 &0.22 &-0.03 &3.03 & 71 &266 \\
81 &                       &                    & 0.1 &\phantom{-}0.3 &...  &...  &...   &...  &  3 & 58 \\
82 &                       &0.3  $< R_e < $ 0.5 &-0.3 &-0.1 &2.38 &0.40 &-0.13 &2.68 & 35 &170 \\
83 &                       &                    &-0.1 &\phantom{-}0.1 &2.39 &0.42 &-0.01 &2.82 &149 &408 \\
84 &                       &                    & 0.1 &\phantom{-}0.3 &2.37 &0.41 &\phantom{-}0.14 &2.97 & 12 &116 \\
85 &                       &0.5  $< R_e < $ 0.7 &-0.3 &-0.1 &2.38 &0.56 &-0.13 &2.49 & 26 &152 \\
86 &                       &                    &-0.1 &\phantom{-}0.1 &2.39 &0.61 &\phantom{-}0.01 &2.61 &242 &527 \\
87 &                       &                    & 0.1 &\phantom{-}0.3 &2.39 &0.63 &\phantom{-}0.13 &2.70 & 43 &244 \\
88 &                       &0.7  $< R_e < $ 0.9 &-0.3 &-0.1 &...  &...  &...   &...  &  3 & 50 \\
89 &                       &                    &-0.1 &\phantom{-}0.1 &2.39 &0.76 &\phantom{-}0.02 &2.42 &153 &420 \\
90 &                       &                    & 0.1 &\phantom{-}0.3 &2.40 &0.77 &\phantom{-}0.12 &2.54 & 26 &190 \\
\enddata
\tablecomments{$I_e$ is computed in the $V$-band. \\ {*} Quoted values are in $\log$ units, i.e., $\log \sigma$, $\log R_e$,
  and $\Delta \log I_e$.  \\ {\dag} Total number of galaxies in the
  stacked spectrum. \\  {\ddag} Effective median $S/N$ of the stacked spectrum.}
\end{deluxetable*}

\clearpage

\LongTables
\begin{deluxetable*}{cc@{ $\pm$ }cc@{ $\pm$ }cc@{ $\pm$ }cr@{ $\pm$ }cr@{ $\pm$ }cr@{ $\pm$ }cr@{ $\pm$ }c}
\tabletypesize{\scriptsize}
\tablecaption{Lick Indices and Stellar Population Properties of
  Stacked Spectra\label{index_tab}}
\tablewidth{0pt}
\tablehead{
\colhead{} &
\multicolumn{2}{c}{H$\beta$} &
\multicolumn{2}{c}{$\langle$Fe$\rangle$} &
\multicolumn{2}{c}{Mg $b$} &
\multicolumn{2}{c}{Age\tablenotemark{\dag}} &
\multicolumn{2}{c}{[Fe/H]} &
\multicolumn{2}{c}{[Mg/H]} &
\multicolumn{2}{c}{[Mg/Fe]} \\
\colhead{} &
\multicolumn{2}{c}{({\AA})} &
\multicolumn{2}{c}{({\AA})} &
\multicolumn{2}{c}{({\AA})} &
\multicolumn{2}{c}{(Gyr)} &
\multicolumn{2}{c}{(dex)} &
\multicolumn{2}{c}{(dex)} &
\multicolumn{2}{c}{(dex)} 
}
\startdata
 1 &1.82 &0.08 &2.11 &0.09 &3.29 &0.08 &10.3 &1.5 &-0.48 &0.05 &-0.22 &0.08 &0.26 &0.06 \\ 
 2 &2.00 &0.04 &2.24 &0.05 &3.10 &0.04 & 7.1 &0.5 &-0.33 &0.03 &-0.18 &0.03 &0.15 &0.02 \\
 3 &2.23 &0.06 &2.50 &0.07 &2.93 &0.06 & 4.1 &0.5 &-0.09 &0.04 &-0.06 &0.05 &0.03 &0.03 \\
 4 &1.81 &0.05 &2.25 &0.05 &3.21 &0.05 &10.1 &0.9 &-0.39 &0.03 &-0.23 &0.03 &0.16 &0.02 \\
 5 &1.99 &0.03 &2.40 &0.03 &3.27 &0.03 & 6.9 &0.3 &-0.22 &0.01 &-0.10 &0.03 &0.12 &0.02 \\
 6 &2.20 &0.03 &2.35 &0.04 &3.02 &0.03 & 4.7 &0.3 &-0.19 &0.02 &-0.09 &0.03 &0.10 &0.02 \\
 7 &1.86 &0.05 &2.36 &0.05 &3.30 &0.04 & 8.9 &0.7 &-0.29 &0.03 &-0.15 &0.04 &0.14 &0.03 \\
 8 &2.07 &0.03 &2.36 &0.03 &3.27 &0.03 & 5.8 &0.3 &-0.23 &0.02 &-0.07 &0.03 &0.16 &0.02 \\
 9 &2.17 &0.05 &2.37 &0.05 &3.37 &0.04 & 4.9 &0.4 &-0.17 &0.03 & 0.02 &0.04 &0.18 &0.03 \\
10 &1.86 &0.09 &2.48 &0.11 &3.36 &0.09 & 8.8 &1.4 &-0.22 &0.05 &-0.14 &0.07 &0.08 &0.05 \\
11 &1.96 &0.05 &2.45 &0.06 &3.55 &0.05 & 7.2 &0.6 &-0.19 &0.03 &-0.03 &0.04 &0.16 &0.02 \\
12 &2.12 &0.08 &2.27 &0.09 &3.13 &0.08 & 5.4 &0.7 &-0.26 &0.05 &-0.09 &0.07 &0.17 &0.05 \\
13 &\ldots  &\ldots  &\ldots  &\ldots  &\ldots  &\ldots  &\ldots  &\ldots &\ldots   &\ldots  &\ldots   &\ldots  &\ldots  &\ldots  \\
14 &1.78 &0.09 &2.51 &0.11 &3.40 &0.09 &10.1 &1.6 &-0.24 &0.05 &-0.15 &0.07 &0.09 &0.05 \\
15 &\ldots  &\ldots  &\ldots  &\ldots  &\ldots  &\ldots  &\ldots  &\ldots &\ldots   &\ldots  &\ldots   &\ldots  &\ldots  &\ldots  \\
16 &1.77 &0.06 &2.49 &0.06 &3.35 &0.06 &10.4 &1.0 &-0.26 &0.03 &-0.18 &0.04 &0.08 &0.03 \\
17 &1.92 &0.03 &2.42 &0.03 &3.31 &0.03 & 7.8 &0.4 &-0.23 &0.02 &-0.12 &0.03 &0.12 &0.02 \\
18 &2.21 &0.04 &2.46 &0.05 &3.19 &0.04 & 4.5 &0.6 &-0.11 &0.03 &-0.02 &0.03 &0.09 &0.02 \\
19 &1.83 &0.04 &2.42 &0.04 &3.56 &0.03 & 9.2 &0.6 &-0.25 &0.02 &-0.08 &0.03 &0.17 &0.02 \\
20 &1.97 &0.02 &2.47 &0.02 &3.53 &0.02 & 6.9 &0.2 &-0.16 &0.01 &-0.01 &0.01 &0.15 &0.01 \\
21 &2.13 &0.03 &2.43 &0.04 &3.42 &0.03 & 5.2 &0.3 &-0.15 &0.02 & 0.01 &0.03 &0.16 &0.02 \\
22 &1.82 &0.03 &2.44 &0.04 &3.41 &0.03 & 9.3 &0.6 &-0.26 &0.02 &-0.13 &0.03 &0.13 &0.02 \\
23 &1.93 &0.02 &2.49 &0.02 &3.57 &0.02 & 7.4 &0.2 &-0.17 &0.01 &-0.02 &0.01 &0.15 &0.01 \\
24 &2.10 &0.03 &2.55 &0.04 &3.49 &0.03 & 5.3 &0.3 &-0.08 &0.02 & 0.05 &0.03 &0.13 &0.02 \\
25 &1.83 &0.06 &2.32 &0.07 &3.44 &0.06 & 9.4 &1.1 &-0.32 &0.04 &-0.13 &0.06 &0.19 &0.05 \\
26 &1.85 &0.03 &2.51 &0.04 &3.59 &0.03 & 8.6 &0.5 &-0.19 &0.02 &-0.05 &0.03 &0.14 &0.02 \\
27 &2.14 &0.06 &2.50 &0.07 &3.53 &0.06 & 4.9 &0.6 &-0.09 &0.04 & 0.07 &0.05 &0.16 &0.03 \\
28 &1.96 &0.12 &2.07 &0.15 &3.28 &0.12 & 7.8 &1.6 &-0.45 &0.09 &-0.13 &0.12 &0.32 &0.09 \\
29 &1.85 &0.09 &2.57 &0.10 &3.57 &0.08 & 8.6 &1.3 &-0.16 &0.05 &-0.04 &0.07 &0.12 &0.05 \\
30 &\ldots  &\ldots  &\ldots  &\ldots  &\ldots  &\ldots  &\ldots  &\ldots &\ldots   &\ldots  &\ldots   &\ldots  &\ldots  &\ldots  \\
31 &1.73 &0.04 &2.46 &0.04 &3.66 &0.04 &11.1 &0.8 &-0.27 &0.02 &-0.10 &0.03 &0.17 &0.02 \\
32 &1.84 &0.02 &2.49 &0.02 &3.66 &0.02 & 8.7 &0.3 &-0.20 &0.01 &-0.04 &0.02 &0.16 &0.02 \\
33 &2.04 &0.03 &2.51 &0.03 &3.51 &0.03 & 6.0 &0.4 &-0.13 &0.02 & 0.01 &0.03 &0.14 &0.02 \\
34 &1.80 &0.03 &2.45 &0.03 &3.79 &0.03 & 9.5 &0.5 &-0.23 &0.02 &-0.02 &0.03 &0.21 &0.02 \\
35 &1.86 &0.01 &2.56 &0.01 &3.79 &0.01 & 8.3 &0.2 &-0.14 &0.01 & 0.03 &0.01 &0.17 &0.01 \\
36 &2.06 &0.02 &2.61 &0.02 &3.58 &0.02 & 5.5 &0.2 &-0.05 &0.01 & 0.07 &0.02 &0.12 &0.01 \\
37 &1.74 &0.03 &2.47 &0.03 &3.85 &0.03 &10.6 &0.5 &-0.24 &0.02 &-0.03 &0.03 &0.21 &0.02 \\
38 &1.84 &0.01 &2.57 &0.01 &3.86 &0.01 & 8.5 &0.2 &-0.13 &0.01 & 0.05 &0.01 &0.18 &0.01 \\
39 &2.04 &0.02 &2.60 &0.02 &3.60 &0.02 & 5.8 &0.3 &-0.06 &0.01 & 0.06 &0.02 &0.13 &0.01 \\
40 &1.76 &0.04 &2.37 &0.05 &3.67 &0.04 &10.6 &0.8 &-0.31 &0.03 &-0.08 &0.04 &0.23 &0.03 \\
41 &1.81 &0.02 &2.58 &0.02 &3.84 &0.02 & 9.1 &0.3 &-0.15 &0.01 & 0.02 &0.01 &0.17 &0.01 \\
42 &1.99 &0.03 &2.65 &0.03 &3.68 &0.03 & 6.4 &0.5 &-0.04 &0.02 & 0.07 &0.03 &0.11 &0.02 \\
43 &1.89 &0.11 &2.23 &0.13 &3.59 &0.11 & 8.5 &1.6 &-0.34 &0.07 &-0.06 &0.09 &0.28 &0.06 \\
44 &1.72 &0.04 &2.59 &0.05 &3.89 &0.04 &10.7 &0.8 &-0.18 &0.02 &-0.01 &0.03 &0.17 &0.02 \\
45 &1.84 &0.05 &2.67 &0.06 &3.91 &0.05 & 8.3 &0.8 &-0.07 &0.03 & 0.08 &0.04 &0.15 &0.03 \\
46 &1.56 &0.04 &2.46 &0.05 &4.00 &0.04 &15.2 &1.7 &-0.19 &0.04 & 0.07 &0.04 &0.26 &0.02 \\
47 &1.75 &0.02 &2.46 &0.02 &3.96 &0.02 &10.2 &0.4 &-0.23 &0.01 & 0.02 &0.02 &0.25 &0.02 \\
48 &1.92 &0.03 &2.61 &0.04 &3.90 &0.03 & 7.3 &0.5 &-0.08 &0.02 & 0.10 &0.03 &0.18 &0.02 \\
49 &1.69 &0.03 &2.48 &0.03 &3.99 &0.02 &11.4 &0.5 &-0.25 &0.01 & 0.01 &0.03 &0.26 &0.03 \\
50 &1.78 &0.01 &2.61 &0.01 &4.08 &0.01 & 9.4 &0.2 &-0.12 &0.01 & 0.08 &0.01 &0.21 &0.01 \\
51 &1.97 &0.02 &2.67 &0.02 &3.85 &0.02 & 6.6 &0.2 &-0.02 &0.01 & 0.12 &0.01 &0.14 &0.01 \\
52 &1.75 &0.03 &2.50 &0.03 &4.13 &0.02 &10.2 &0.5 &-0.20 &0.01 & 0.07 &0.03 &0.27 &0.02 \\
53 &1.76 &0.01 &2.65 &0.01 &4.10 &0.01 & 9.6 &0.2 &-0.11 &0.01 & 0.08 &0.02 &0.19 &0.02 \\
54 &1.94 &0.02 &2.67 &0.02 &3.91 &0.01 & 7.0 &0.2 &-0.03 &0.01 & 0.13 &0.02 &0.16 &0.02 \\
55 &1.75 &0.04 &2.42 &0.04 &4.20 &0.04 &10.4 &0.7 &-0.24 &0.02 & 0.08 &0.03 &0.32 &0.02 \\
56 &1.77 &0.01 &2.66 &0.01 &4.06 &0.01 & 9.4 &0.2 &-0.10 &0.01 & 0.08 &0.01 &0.18 &0.01 \\
57 &1.90 &0.02 &2.63 &0.02 &3.91 &0.02 & 7.5 &0.3 &-0.07 &0.01 & 0.10 &0.02 &0.17 &0.01 \\
58 &1.77 &0.07 &2.34 &0.08 &4.15 &0.07 &10.1 &1.3 &-0.29 &0.04 & 0.07 &0.06 &0.36 &0.05 \\
59 &1.70 &0.02 &2.62 &0.03 &4.05 &0.02 &10.9 &0.5 &-0.16 &0.01 & 0.03 &0.02 &0.19 &0.02 \\
60 &1.83 &0.04 &2.57 &0.04 &4.05 &0.04 & 8.5 &0.6 &-0.12 &0.02 & 0.11 &0.03 &0.23 &0.02 \\
61 &1.59 &0.05 &2.44 &0.06 &4.35 &0.05 &13.8 &1.3 &-0.28 &0.03 & 0.05 &0.05 &0.33 &0.04 \\
62 &1.65 &0.03 &2.59 &0.03 &4.40 &0.03 &11.8 &0.6 &-0.17 &0.02 & 0.11 &0.03 &0.28 &0.02 \\
63 &1.76 &0.07 &2.73 &0.09 &4.25 &0.07 & 9.3 &1.1 &-0.05 &0.04 & 0.14 &0.06 &0.19 &0.05 \\
64 &1.60 &0.03 &2.59 &0.04 &4.32 &0.03 &12.9 &0.7 &-0.19 &0.02 & 0.07 &0.03 &0.26 &0.02 \\
65 &1.66 &0.02 &2.61 &0.02 &4.40 &0.02 &11.4 &0.3 &-0.15 &0.01 & 0.13 &0.01 &0.28 &0.01 \\
66 &1.82 &0.03 &2.79 &0.04 &4.19 &0.03 & 8.4 &0.5 & 0.01 &0.01 & 0.17 &0.03 &0.16 &0.02 \\
67 &1.58 &0.03 &2.59 &0.03 &4.44 &0.03 &13.3 &0.7 &-0.19 &0.02 & 0.09 &0.03 &0.28 &0.02 \\
68 &1.70 &0.01 &2.69 &0.01 &4.34 &0.01 &10.5 &0.2 &-0.09 &0.01 & 0.14 &0.01 &0.23 &0.01 \\
69 &1.88 &0.02 &2.73 &0.02 &4.13 &0.02 & 7.5 &0.3 &-0.00 &0.01 & 0.18 &0.02 &0.18 &0.02 \\
70 &1.60 &0.04 &2.58 &0.04 &4.47 &0.04 &12.9 &0.9 &-0.19 &0.02 & 0.11 &0.03 &0.30 &0.02 \\
71 &1.69 &0.01 &2.69 &0.01 &4.39 &0.01 &10.7 &0.2 &-0.10 &0.01 & 0.14 &0.02 &0.24 &0.02 \\
72 &1.87 &0.02 &2.76 &0.02 &4.20 &0.02 & 7.6 &0.2 & 0.01 &0.01 & 0.19 &0.01 &0.18 &0.01 \\
73 &1.60 &0.08 &2.46 &0.09 &4.68 &0.07 &13.1 &1.8 &-0.24 &0.05 & 0.17 &0.08 &0.41 &0.06 \\
74 &1.65 &0.02 &2.67 &0.02 &4.35 &0.02 &11.5 &0.3 &-0.12 &0.01 & 0.11 &0.02 &0.24 &0.02 \\
75 &1.84 &0.02 &2.74 &0.03 &3.22 &0.02 & 8.2 &0.4 &-0.05 &0.01 & 0.03 &0.82 &0.08 &0.82 \\
76 &1.47 &0.08 &2.50 &0.10 &4.59 &0.08 &\ldots &\ldots\tablenotemark{\dag} &-0.19 &0.06 & 0.20 &0.08 &0.38 &0.05 \\
77 &1.62 &0.04 &2.71 &0.05 &4.58 &0.04 &11.8 &0.8 &-0.09 &0.03 & 0.18 &0.04 &0.27 &0.03 \\
78 &\ldots  &\ldots  &\ldots  &\ldots  &\ldots  &\ldots  &\ldots  &\ldots &\ldots   &\ldots  &\ldots   &\ldots  &\ldots  &\ldots  \\
79 &1.53 &0.05 &2.68 &0.06 &4.72 &0.05 &14.0 &1.4 &-0.14 &0.03 & 0.16 &0.05 &0.30 &0.04 \\
80 &1.62 &0.03 &2.75 &0.03 &4.60 &0.03 &11.7 &0.5 &-0.07 &0.02 & 0.19 &0.03 &0.26 &0.03 \\
81 &\ldots  &\ldots  &\ldots  &\ldots  &\ldots  &\ldots  &\ldots  &\ldots &\ldots   &\ldots  &\ldots   &\ldots  &\ldots  &\ldots  \\
82 &1.55 &0.04 &2.55 &0.05 &4.67 &0.04 &13.7 &1.1 &-0.20 &0.03 & 0.17 &0.04 &0.37 &0.04 \\
83 &1.58 &0.02 &2.74 &0.02 &4.67 &0.02 &12.4 &0.3 &-0.09 &0.01 & 0.19 &0.01 &0.28 &0.01 \\
84 &1.79 &0.07 &2.87 &0.08 &4.60 &0.06 & 8.4 &1.0 & 0.05 &0.03 & 0.27 &0.04 &0.22 &0.03 \\
85 &1.55 &0.05 &2.66 &0.05 &4.63 &0.05 &13.5 &1.2 &-0.15 &0.03 & 0.15 &0.04 &0.30 &0.02 \\
86 &1.63 &0.01 &2.76 &0.02 &4.57 &0.01 &11.5 &0.3 &-0.06 &0.01 & 0.19 &0.01 &0.25 &0.01 \\
87 &1.71 &0.03 &2.80 &0.03 &4.47 &0.03 & 9.9 &0.5 &-0.02 &0.02 & 0.19 &0.03 &0.21 &0.02 \\
88 &\ldots  &\ldots  &\ldots  &\ldots  &\ldots  &\ldots  &\ldots  &\ldots &\ldots   &\ldots  &\ldots   &\ldots  &\ldots  &\ldots  \\
89 &1.60 &0.02 &2.80 &0.02 &4.64 &0.02 &11.9 &0.3 &-0.05 &0.01 & 0.20 &0.01 &0.25 &0.01 \\
90 &1.74 &0.04 &2.83 &0.04 &4.36 &0.03 & 9.4 &0.6 & 0.01 &0.01 & 0.19 &0.03 &0.18 &0.02 \\
\enddata
\tablecomments{{\dag} Ages are determined from H$\beta$ for all stacked
  spectra with the exception of spectrum 76, whose weak H$\beta$
  absorption falls outside the parameter space covered by the models.
  For this index, H$\gamma_F$ is used in the fitting process to
  determine abundances.  For consistency, the age measured for this
  spectrum is not used in the analysis here, as ages derived from
  H$\gamma_F$ are typically somewhat younger than those derived from
  H$\beta$ (see \citealt{graves08}).}
\end{deluxetable*}

\clearpage

\end{document}